\documentclass[sigconf,screen,nonacm]{acmart}

\usepackage{multirow}
\usepackage{booktabs}
\usepackage{tcolorbox}
\usepackage{amsmath}
\usepackage{xcolor}
\usepackage{threeparttable}
\begin{document}

\title{FastLog: An End-to-End Method to Efficiently Generate and Insert Logging Statements}

\author{Xiaoyuan Xie}
\authornote{Xiaoyuan Xie and Songqiang Chen are the co-corresponding authors.}
\email{xxie@whu.edu.cn}
\affiliation{\institution{{School of Computer Science, Wuhan University}\country{China}}}

\author{Zhipeng Cai}
\email{zhipengcai@whu.edu.cn}
\affiliation{\institution{{School of Computer Science, Wuhan University}\country{China}}}

\author{Songqiang Chen}
\authornotemark[1]
\email{i9s.chen@connect.ust.hk}
\affiliation{\institution{{Department of Computer Science and Engineering, The Hong Kong University of Science and Technology}\country{China}}}
\affiliation{\institution{{School of Computer Science, Wuhan University}\country{China}}}

\author{Jifeng Xuan}
\email{jxuan@whu.edu.cn}
\affiliation{\institution{{School of Computer Science, Wuhan University}\country{China}}}

\begin{abstract}

Logs play a crucial role in modern software systems, serving as a means for developers to record essential information for future software maintenance. As the performance of these log-based maintenance tasks heavily relies on the quality of logging statements, various works have been proposed to assist developers in writing appropriate logging statements. 
However, these works either only support developers in partial sub-tasks of this whole activity; or perform with a relatively high time cost and may introduce unwanted modifications. To address their limitations, we propose FastLog, which can support the complete logging statement generation and insertion activity, in a very speedy manner. Specifically, given a program method, FastLog first predicts the insertion position in the finest token level, and then generates a complete logging statement to insert. We further use text splitting for long input texts to improve the accuracy of predicting where to insert logging statements. 
A comprehensive empirical analysis shows that our method outperforms the state-of-the-art approach in both efficiency and output quality, which reveals its great potential and practicality in current real-time intelligent development environments.

\end{abstract}



\maketitle

\nocite{DATARELEASE}
\section{Introduction}\label{sec_introduction}

Logs have acquired growing significance in modern software systems. They support a wide variety of software maintenance tasks, e.g., testing \cite{testing_chen2017analytics,testing_chen2018automated}, debugging \cite{debugging_satyanarayanan1992transparent}, diagnosis \cite{diagnosis_yuan2010sherlog, diagnosis_yuan2012improving, diagnosis_zhou2019latent}, monitoring \cite{where_statistic_yao2018log4perf, monitoring_hasselbring2020kieker, monitoring_harty2021logging}, anomaly detection \cite{anomaly_du2017deeplog, anomaly_zhang2019robust, anomaly_ludetecting, anomaly_meng2019loganomaly, anomaly_yang2021semi}, and transaction processing \cite{logging_zhou2023scalable}. 
Developers insert logging statements into the source code to record essential information for the future maintenance of software systems. For instance, analyzing log sequences can help to identify anomalies in a software system \cite{anomaly_du2017deeplog}, and matching the execution logs with their corresponding code paths can help to estimate code coverage measures for software testing \cite{testing_chen2018automated}. The effectiveness of these maintenance tasks heavily relies on the quality of the collected logs. Therefore, the significance of writing appropriate logging statements cannot be overstated.

However, writing appropriate logging statements is not an easy task for developers. 
Previous studies suggest that developers often struggle with determining proper logging locations \cite{challenge_where_fu2014developers, where_zhu2015learning, challenge_where_zeng2019studying}, levels \cite{challenge_level_li2020qualitative, challenge_level_li2017log}, and messages \cite{what_yuan2012characterizing}, which are \textbf{three essential sub-tasks in a complete logging statement generation activity}.

To assist developers in writing logging statements more effectively, a variety of approaches have been proposed to assist with different aspects of this process. 
On one hand, several studies have focused on helping developers determine ``where to log'' \cite{where_candido2021exploratory, where_li2018studying, where_statistic_yao2018log4perf, where_zhu2015learning, where_jia2018smartlog, where_li2020shall, where_zhang2023deeplog}. 
Researchers leverage topic modeling and deep learning to suggest logging locations at method level \cite{where_li2018studying} and code-block level \cite{where_li2020shall}, respectively. 
On the other hand, some methods help developers decide ``what to log'' \cite{what_liu2022tell, what_liu2019variables}. Li et al. \cite{what_li2021deeplv} leverage the ordinal nature of log levels and design neural networks to suggest log levels. Ding et al. \cite{what_ding2022logentext} build a neural machine translation model to generate the log messages. 
However, these works only consider tackling one of the aforementioned three sub-tasks. They cannot fully support developers in a complete logging statement generation activity.

Recently, Mastropaolo et al. proposed LANCE \cite{what_mastropaolo2022using}, which is the first end-to-end solution that supports developers in all decisions of logging locations, log levels, and log message content.
Specifically, LANCE employs the T5 \cite{t5_raffel2020exploring} model to generate complete logging statements, taking a program method as input and inserting a complete logging statement in it as output. With the Seq2Seq model, LANCE automates the logging activities in all three aforementioned sub-tasks in a single step, providing comprehensive support to developers in writing logging statements.

Nonetheless, LANCE still suffers from two limitations. Firstly, when regenerating the entire program method, \textbf{there is a risk of unintentionally modifying the non-log code}. This deviates from the requirement of preserving the content other than the inserted logging statement. Secondly and more importantly, due to the need of generating the entire program method token by token, the generation speed of LANCE is fairly low. \textbf{This should greatly hinder this approach from being widely adopted in practice}.

Therefore, in this paper, we propose FastLog, \textbf{which properly addresses the above two limitations while preserving the same end-to-end generation paradigm as LANCE.} 
FastLog can also support developers in all three sub-tasks (i.e., determining logging locations, levels, and messages). It achieves the generation in a two-stage manner.
First, it employs token classification to predict where to insert logging statements at the finest-grained token level, where a text splitting strategy is introduced to enhance its prediction accuracy for lengthy input examples. 
Then, it employs a Seq2Seq model to generate the content of the complete logging statements that need to be inserted. 
As there is no need to regenerate the entire program method, this method enables fast logging statement generation and will not modify the content other than the inserted logging statement. Moreover, by decomposing the complete logging statement generation task into two simpler steps, FastLog tends to reduce the model burden and may contribute to better performance.

\textbf{As a reminder, such a two-stage method is by no means a simple binding of previous methods for each particular sub-task of generation.}
Firstly, previous methods for determining ``where to log'' are at the method or code-block level, which are too coarse-grained to indicate the precise position for inserting the logging statements. We need a new method to pinpoint the location that can be directly adopted by the following generation step to insert the logging statements. 
Secondly, current methods for ``what to log'' usually focus on a particular component of the logging statement (i.e., log level \cite{what_li2021deeplv, what_liu2022tell}, static description text \cite{what_he2018characterizing, what_ding2022logentext}, and variables \cite{what_liu2019variables}), rather than the complete content. Actually, it is not a good practice to first generate content for each component separately with different approaches and then combine their outputs to form the complete one, since this process is low in efficiency, and more importantly, it is very likely to introduce incompatibility among different component outputs.
Therefore, it is necessary to generate the complete logging statement at once to ensure the content compatibility, as well as to increase the efficiency.

We perform extensive experiments to evaluate the performance of the proposed method. Besides using a benchmark adopted in the previous work \cite{what_mastropaolo2022using}, we also build a new dataset to avoid potential data leakage.
The evaluation results on both benchmarks demonstrate that our method contributes to a satisfying improvement in terms of both generation speed and quality. Specifically, our method boosts the generation of complete logging statements with a significant speedup of about 12 times compared to LANCE (0.22s v.s. 2.80s per sample). Meanwhile, in comparison to the state-of-the-art method, our method obtains an approximate 5\% accuracy improvement in predicting logging position and log level, and increases the BLEU metric by over 3 and the ROUGE-L metric by around 5 for log messages.

This work makes the following contributions to the domain of logging statement generation.

\begin{itemize}

\item We propose a new end-to-end logging statement generation method called FastLog. In comparison to the peered approach LANCE, FastLog can also support developers in all three essential sub-tasks. Meanwhile, it avoids the risk of modifying contents other than the inserted logging statement in LANCE. More importantly, it significantly reduces the time cost, making it more practical to fit in a real-time intelligent development environment.

\item As compared with the previous location prediction works, FastLog locates the logging positions at the finest-grained token level and leverages text splitting for improving the prediction accuracy. Besides, unlike previous works that only generate partial logging statement content, FastLog generates the complete logging statements to be inserted.

\item We performed a comprehensive empirical evaluation on a previous dataset and a self-constructed dataset. Results show that our method outperforms the state-of-the-art approach in both efficiency and output quality.
We also explored the feasibility of providing multiple options to further enhance the practicality of FastLog. Preliminary experiment shows very promising results.

\end{itemize}

The rest of the paper is organized as follows. Section~\ref{sec_motivation} and \ref{sec_challenges} describe the motivations and challenges of our work, respectively. Section~\ref{sec_methodology} elaborates our proposed method, FastLog. Section~\ref{sec_setup}, \ref{sec_results}, and \ref{sec_discussion} present the setting and analyze the results of our evaluation experiments. Section~\ref{sec_threats} discusses the potential threats to validity. Section~\ref{sec_relatedworks} lists some related works. Finally, Section~\ref{sec_conclusion} gives a conclusion and lists our future work.

\section{Motivation}\label{sec_motivation}

\begin{figure}[tbp]
\centerline{\includegraphics[width=0.996\linewidth]{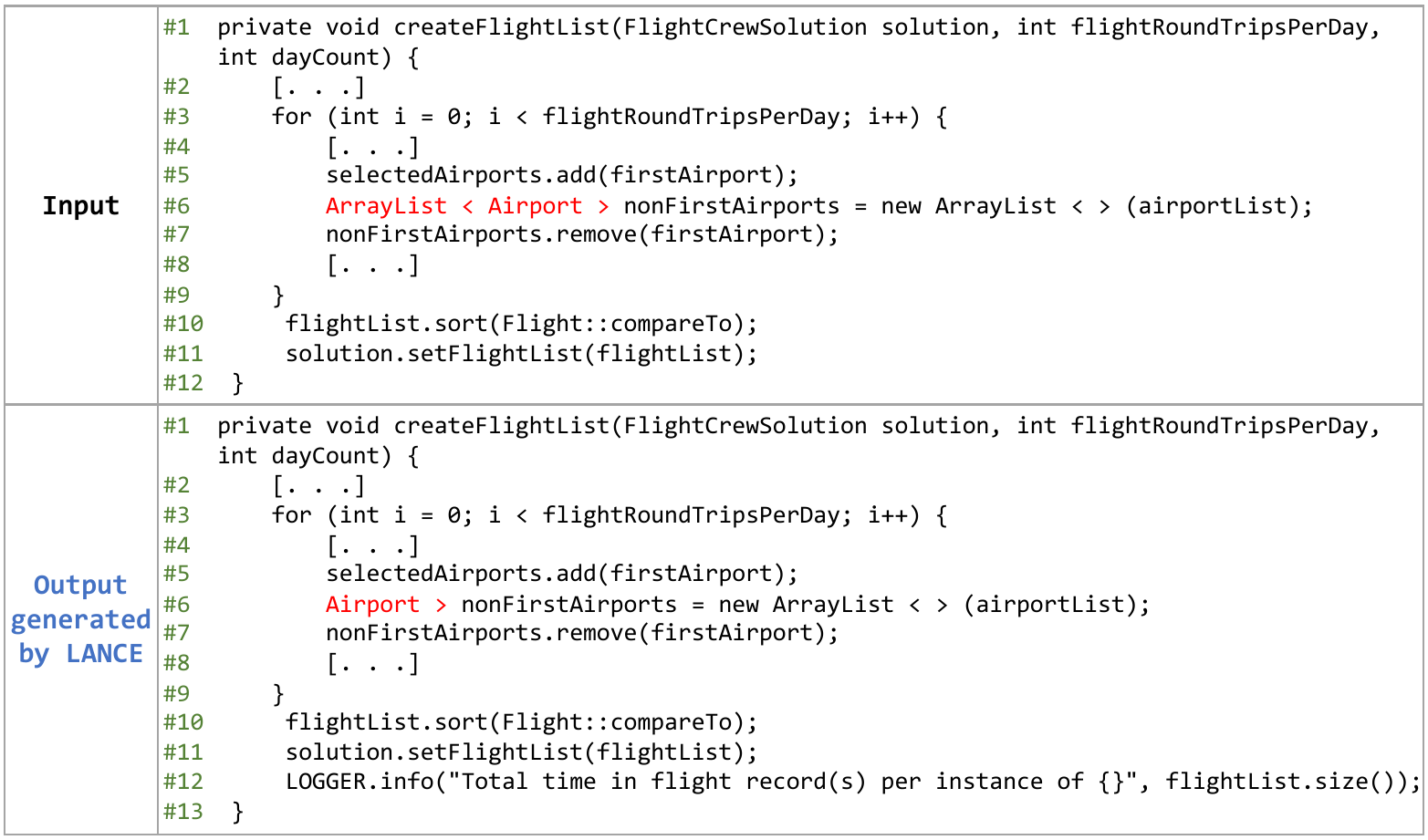}}
\caption{An Example Output Generated by LANCE}
\label{motivation_example}
\end{figure}

LANCE \cite{what_mastropaolo2022using} is the state-of-the-art method to concurrently automate logging activities in three essential aspects: determining the logging position, selecting the appropriate verbosity log level, and specifying the log message content. It utilizes a Seq2Seq approach to generate and insert a complete logging statement into the given program method. LANCE takes a program method that lacks a logging statement as the input and outputs a whole program method with a newly inserted logging statement at the appropriate location. However, this approach has the following two limitations:

\textbf{(1) Potential risk for modifying the context content other than the inserted logging statement.} The text generation process of a Seq2Seq model involves using an iterative decoding algorithm to select the next token based on a probability distribution, without directly copying content existing in the input text. This mechanism introduces an element of uncertainty in the generation process \cite{uncertainty_giulianelli2023comes}. Thus, LANCE may inadvertently modify the text other than the inserted logging statement when regenerating the whole program method. Upon inspecting the outputs generated by LANCE, we observed some such cases where the final output included modifications to text other than the inserted logging statement. For example, in the case in Fig.~\ref{motivation_example}, LANCE inadvertently altered the term ``{{ArrayList \textless Airport\textgreater}}'' to ``{{Airport\textgreater}}'' when regenerating the non-log code. This deviates from the desired outcome of inserting a logging statement while preserving the remaining content other than the inserted logging statement.
We have summarized two major categories of unwanted modifications by LANCE and give more examples in Section~\ref{subsec_modificationcases}.

\textbf{(2) Inefficiency in generating and inserting complete logging statements.} Another more important limitation of LANCE is its inefficiency in generating and inserting complete logging statements. As previously explained, LANCE is designed to output a modified program method with the inserted logging statement. This means that LANCE needs to regenerate the content of the entire input program method. Unfortunately, text generation, particularly for lengthy texts, is time-consuming and resource-intensive \cite{long_liang2023open}. Therefore, it can be inefficient to generate and insert a logging statement for a given program method in this way. 
The second row of Fig.~\ref{motivation_example} illustrates an example output generated by LANCE. It can be found that apart from the real logging statement on code line 12, all the other contents are also regenerated together. Actually, \textbf{regenerating the content other than the real logging statement is unnecessary and incurs inefficiency}. According to our statistics, even on a workstation with a premium GPU NVIDIA GeForce RTX 3090 with an inference batch size of 1, LANCE requires an average of 2.8 seconds to generate a single sample result.  
However, based on our assessment, about 92\% of the tokens in the original test dataset \textbf{do not belong to the target logging statements}, which leads to a significant waste of time. This demonstrates the limited practicality of LANCE to fit in a real-time intelligent development environment that always requires instant feedback.

To address the aforementioned issues, we aim to explore an efficient and enhanced method to accomplish the generation and insertion of complete logging statements. 
To avoid modifying the content other than the inserted logging statement and enable an efficient generation process, we can only focus on generating the needed logging statement, rather than the entire program method. 
Besides, we need another step to first locate the logging position where a logging statement is required. Thus, to accomplish the entire task, we need two steps, i.e., locating the logging position and then generating the needed logging statement to be inserted.

\section{Challenges}\label{sec_challenges}

By revisiting the pipeline of the logging statement generation task, we reckon that as compared with LANCE which regenerates a whole program method, \textbf{dividing the entire task into sub-tasks is still a reasonable manner}. However, except LANCE, existing works all focus on one particular sub-task and there is no work that properly links all the sub-tasks together.
In fact, such a pipeline is by no means a simple binding of previous methods for each particular step of generation. We face the following challenges.

\textbf{(1) Achieving efficient and fine-grained logging position prediction.} 

Existing works for ``where to log'' predict whether a logging statement is needed within a given method \cite{where_li2018studying, where_candido2021exploratory} or a given code block (that includes several statements) \cite{where_zhu2015learning,  where_jia2018smartlog, where_li2020shall, where_zhang2023deeplog}, where the prediction is either ``yes'' or ``no''.
Such a coarse result cannot indicate the precise position to insert the logging statement.

In our task, in order to fit in the complete pipeline, we need to pinpoint the precise location of the inserted logging statement \textbf{on the token level} (i.e., determining the token after which a logging statement should be inserted). Given a code snippet, there are lots of tokens to choose from, leading to a complex decision process rather than a simple binary choice, with multiple options related to the number of tokens.

\textbf{(2) Generating an entire high-quality logging statement with limited information.} 

Existing methods for ``what to log'' usually focus on a particular component of the logging statement (i.e., log level \cite{what_li2021deeplv, what_liu2022tell}, static description text \cite{what_he2018characterizing, what_ding2022logentext}, and variables \cite{what_liu2019variables}), rather than the complete content. 
However, simply binding the outcomes of different components may not be a feasible solution. One challenge is to guarantee that these components from separate models are compatible with each other. For example, if the predicted log level is ``Error'' but the generated text from another model is related to a calculated metric like consumed time, then they are not describing the same thing and hence binding them together does not make any sense.

Therefore, in order to avoid such incompatibility, we need to generate the complete logging statement at once. However, this means we can utilize less information when generating logging statement content. As an example, Li et al. \cite{what_li2021deeplv} combine AST information, which inherently indicates logging positions, with log messages to better predict log levels. However, since we predict log levels by generating complete logging statements, we are unable to leverage the log messages as input information to enhance our predictions. We can only use limited information from the input program method and the predicted insertion position from the first step to generate the complete logging statement.

\begin{figure*}[ht]
\centerline{\includegraphics[width=1.0\textwidth]
{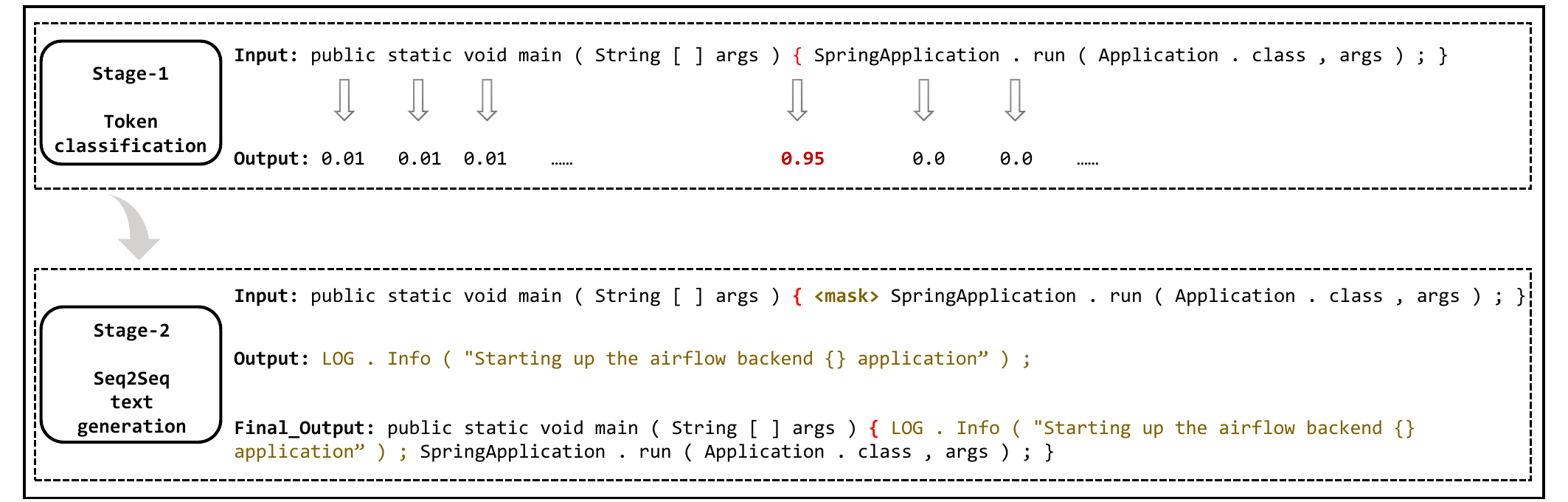}}
\caption{Overview of Our Method}
\label{method_overview}
\end{figure*}

\section{Methodology}\label{sec_methodology}

\subsection{Overview}

Fig.~\ref{method_overview} illustrates the overall process of our method FastLog, which aims to generate and insert complete logging statements via two stages. Firstly, FastLog employs token classification to predict the insertion position of a needed logging statement. Then, it utilizes a Seq2Seq model to generate the content of the logging statement to be inserted.

In Stage-1, unlike previous works that decide the method or code block that needs log insertion, we need to decide the appropriate token after which the logging statement should be inserted.

To achieve this goal, we adopt token classification to predict whether a logging statement is needed after a particular token.
Specifically, each token in the input text is subjected to binary classification, where a label ``1'' indicates the need to insert a logging statement after that token. We obtain the probability of each token being classified as ``1'' and select the token with the highest probability as the insertion position for the logging statement.

In Stage-2, based on the predicted insertion position from Stage-1, we insert a ``$<$mask$>$'' token as a placeholder in the input text to indicate the task of generating a logging statement for that position. Subsequently, we utilize a Seq2Seq model to generate the content of the logging statement, including all components like log level, static text, and involved variables. Upon obtaining the content of the logging statement, we replace the ``$<$mask$>$'' token in the input text with the generated logging statement, resulting in the final output, a program method with an inserted logging statement. 

To summarize, even though FastLog acts in a classical pipeline for logging statement generation, it is actually \textbf{the first and only work that can truly link each stage together to accomplish the entire task}, by tackling the aforementioned challenges in Section~\ref{sec_challenges}. 
And as compared with the peer work LANCE (in a different pipeline): (1) pinpointing the insertion position enables us to focus on generating the logging statement only, as well as avoids inefficient regeneration of the entire method and any unintentional modification to the non-log statements; while (2) generating and inserting the complete logging statement further accomplishes the task. \textbf{Thus, FastLog can intuitively address the two limitations of LANCE} that we discuss in Section~\ref{sec_motivation}.

In the implementation, we utilize PLBART \cite{plbart_ahmad2021unified} as the base model to fine-tune two separate models for Stage-1 and Stage-2, respectively.

\subsection{Stage-1: Predicting Insertion Position}
In Stage-1, we utilize token classification to predict the insertion positions of the required logging statements. Let \(X\) be the set of tokens in the input text. During the training phase, we conduct binary classification for each token \(x_i \in X\) in the input text. Let \(y_i\) represent the label for token \(x_i\), where \(y_i = 1\) indicates the need to insert a logging statement after token \(x_i\) and \(y_i = 0\) indicates that no logging statement is needed after token \(x_i\). Therefore, we have a set of labeled examples \(\{(x_i, y_i)\}_{i=1}^{|X|}\) for training.

In the inference phase, since our goal is to identify the insertion point for the required logging statement, we only collect the probabilities of each token being classified as ``1''. Specifically, given an input text with tokens \(X\), we collect the probability \(P(y_i=1|x_i)\) for each token \(x_i \in X\). To determine the insertion point, we select the token \(x_{\text{insert}}\) with the highest probability, i.e., \(x_{\text{insert}} = \arg\max_{x_i \in X} P(y_i=1|x_i)\). Then a logging statement should be inserted after the token \(x_{\text{insert}}\).
Besides, through the syntactic analysis, we can narrow down the selection to only include the tokens ``\{'', ``\}'', ``;'' and ``:'', since logging statements only occur after these four tokens.

Ideally, the token classification task should classify every token in the input text. However, due to the inherent limitations of Natural Language Processing (NLP) models, there exists a maximum length constraint for the model input. When the length of the input text exceeds this limit, the exceeding portion will be truncated and discarded. 
Inevitably, there will be some lengthy code snippets in the dataset. As a result, these long samples do not obtain complete classification results for all tokens during the prediction process.

To obtain complete token classification results for the input text during the inference phase, we design a strategy of splitting lengthy input text into smaller chunks. For the lengthy input text, we divide it into multiple chunks. Each chunk will be individually fed into the model for prediction. Subsequently, we combine the classification results from these chunks to formulate a comprehensive final result.
Our splitting strategy is as follows: 
For the input codes whose length does not exceed the maximum model input length threshold (denoted as \textit{$\text{max\_model\_input\_length}$}), we do not split them.
Meanwhile, for the input codes that exceed \textit{$\text{max\_model\_input\_length}$}, we evenly split them into multiple chunks, with each chunk not exceeding the maximum length for a chunk (denoted as \textit{$\text{max\_chunk\_length}$}). 
We propose this even splitting strategy to avoid generating too short chunks, which will cause significant semantic loss and negatively impact the model prediction.

Besides, it is worth noting that dividing long input texts into chunks may result in missing context for tokens located near the boundaries of each chunk due to truncation. To address this problem, we propose to add a set of contextual tokens to the truncated boundaries of each chunk to compensate for the missing context.
Specifically, we include certain contextual tokens that originally occupied those positions at the truncated boundaries of each chunk. 
To offer fairly complete contextual semantics, we also ensure that each chunk consists of complete statements, where a complete statement should end with ``;'' or ``\}''.
We add a specific number (denoted as \textit{k}) of statements to the truncated boundaries of each chunk. 
As a reminder, in the implementation, we also set a constraint to ensure that the length of each chunk after adding contextual tokens does not exceed \textit{$\text{max\_model\_input\_length}$}.
Specifically, the contextual tokens added to the truncated boundaries of each chunk, on both the left and right sides if necessary, should be no more than $\frac{1}{2}\left(\text{max\_model\_input\_length} - \text{max\_chunk\_length}\right)$.
Thus, the actual number of complete statements to be added may be less than \textit{k}.
In addition, we do not consider the prediction results for the added assistant contextual tokens in each chunk.

Finally, for all model inputs (including non-split input and split input), we utilize padding and truncation techniques to ensure that the token length of each input is equal to \textit{$\text{max\_model\_input\_length}$}.

\subsection{Stage-2: Generating Logging Statement}

In Stage-2, we employ the Seq2Seq text generation approach to generate the content (including log levels and log messages) of logging statements. In previous work on logging text generation, the model input consisted of the code preceding the target logging statement \cite{what_ding2022logentext}. This provided explicit information to the model about the location of the logging statements.  
Recognizing that the insertion positions of logging statements may help generate their contents, we also try to provide this information for the model in our method.

Thus, in our method, the input to the Seq2Seq model is not the original program method, but a modified text inserted with a special ``$<$mask$>$'' token. The ``$<$mask$>$'' token serves as a placeholder to indicate the position where the logging statement is required to be inserted. Its specific position is determined by the prediction result from Stage-1.

Let \(X'\) be the set of tokens in the modified input code. To facilitate the Seq2Seq model to generate the logging statement, we insert a ``\textless mask\textgreater'' placeholder after the token \(x_{\text{insert}}\) determined in Stage-1. Thus, the input of Stage-2 can be denoted as \(X' = \{x_1, x_2, \ldots, x_{\text{insert}}, \text{\textless mask\textgreater}, x_{\text{insert}+1}, \ldots, x_{|X'|}\}\). By providing this modified input text \(X'\), we guide the model to focus on the context surrounding the insertion position.

For the input text that exceeds \textit{$\text{max\_model\_input\_length}$}, we employ the same splitting strategy in Stage-1 to split the input text and use the chunk containing the ``$<$mask$>$'' token as the model input. This ensures that the input provided to the model effectively conveys the positional information about where a logging statement is needed.

\subsection{Fine-tuning PLBART}

We choose the PLBART \cite{plbart_ahmad2021unified} model as the foundation for our method and fine-tune two separate models to handle the token classification and Seq2Seq text generation tasks for two stages, respectively.
This choice is motivated by the following reasons: In our preliminary experiments, we have tried several mainstream models. We found that the token classification task did not yield satisfactory results with models from the T5 \cite{t5_raffel2020exploring} series, while the text generation task did not perform well with models from the BERT \cite{bert_devlin2018bert} series. Meanwhile, the BART \cite{bart_lewis2019bart} series models, including PLBART, demonstrate strong performance in both tasks. Additionally, the PLBART model has been pre-trained on code corpora, making it a suitable choice for our method to generate and insert complete logging statements.

\section{Evaluation Setup}\label{sec_setup}

We evaluate our proposed method with the following four research questions:

\textbf{RQ1: How does FastLog compare to the state-of-the-art approach in terms of output quality?} This RQ aims to evaluate the effectiveness of our method in predicting accurate logging positions and statement content.

\textbf{RQ2: How efficient is FastLog compared to the state-of-the-art approach?} This RQ aims to evaluate the efficiency of our method in providing complete decisions on logging locations, log levels, and log messages.

\textbf{RQ3: Beyond the overall accuracy, how well does FastLog predict logging positions and levels from a more granular perspective?} This RQ aims to analyze the characteristics of all prediction results (including wrong predictions) in detail to further verify the effectiveness of our method.

\textbf{RQ4: Does splitting long input text into smaller chunks benefit the generation and insertion of logging statements?} This RQ aims to study the need of splitting long input text as well as the advantages of our splitting strategies.

\subsection{Dataset}

In this study, we utilize the dataset adopted in the previous work \cite{what_mastropaolo2022using} for training our model and conducting experiments. 
However, both this dataset and the pre-training sample set for the PLBART model draw data from GitHub repositories, and they exhibit a temporal overlap. Consequently, certain samples in the test dataset used in \cite{what_mastropaolo2022using} might have been encountered by the PLBART model during its pre-training phase. 
To avoid making any data leakage and introducing bias in evaluation results, we construct a new test dataset to ensure that all test samples are completely new to the model. The process of constructing this new test dataset follows the previous work. Specifically, we utilized the search tool provided by Dabic et al. \cite{dataset_dabic2021sampling} to retrieve recent GitHub repositories that match the following characteristics:
\begin{itemize}
  \item The repository was created between September 1, 2021, and May 1, 2023.
  \item It is a Java language repository.
  \item The project incorporates a dependency on Apache Log4j.
  \item It is not a forked project.
  \item The repository has received at least 10 stars.
\end{itemize}

The details of all datasets used in our study are listed in Table~\ref{dataset_details}. We conducted experiments on both the original test dataset and the new test dataset we constructed, with the original test dataset containing 12,020 test samples, and the new test dataset containing 7,235 test samples. This allows us to better evaluate the effectiveness of our PLBART-based method and confirm that the observed outcomes are not attributed to data leakage.

\begin{table}[]
\caption{The Dataset Details.}
\label{dataset_details}
\resizebox{0.48\textwidth}{!}{
\begin{tabular}{@{}cccc@{}}
\toprule
Dataset &
  Sample Count &
  \begin{tabular}[c]{@{}c@{}}Mean Token Length\\  of Input\end{tabular} &
  \begin{tabular}[c]{@{}c@{}}Mean Token Length of \\ Target Logging Statement\end{tabular} \\ \midrule
Train         & 102036 & 237 & 22 \\
Valid         & 12749  & 236 & 22 \\
Original Test & 12020  & 237 & 21 \\
New Test      & 7235   & 246 & 23 \\ \bottomrule
\end{tabular}}
\end{table}

\subsection{Hyperparameter Settings}

In our experiments, we carefully select the hyperparameters for our model. We set the training batch size as 8, the learning rate as \(2 \times 10^{-5}\), and the weight decay as 0.01. For the logging location prediction task in Stage-1, we trained the model for 10 epochs. For the logging statement generation task in Stage-2, we extended the training to 30 epochs to ensure that the model could better handle the text generation task. 
We select the optimal model checkpoints that perform best on the validation dataset as our final model, where the classification model for Stage 1 and the generation model for Stage-2 are selected according to the F1-score and generation accuracy, respectively. 
For the inference in Stage-2 to generate logging statements, we employ the beam search decoding strategy with a beam size of 10 to predict the output text.

\subsection{Baselines}

As described above, FastLog targets the end-to-end logging statement generation. Among all existing related approaches, only LANCE \cite{what_mastropaolo2022using} can achieve the same goal. Therefore, we select LANCE as our \textbf{major baseline}. The original LANCE is configured with T5\textsubscript{small} model for content generation, so in our comparison, we denote this original LANCE as \textbf{LANCE\textsubscript{T5}}.

Apart from LANCE\textsubscript{T5}, we introduce a variant of LANCE as our \textbf{second baseline}, which is configured with the same generation model (i.e. PLBART) as ours. This is because we want to eliminate the impact of different models and conduct a more fair comparison. We denote this baseline as \textbf{LANCE\textsubscript{PL}}.

As a reminder, both baselines adopt the greedy search decoding strategy. Therefore, to fairly evaluate the effectiveness of our methodology independently of different decoding strategies, we further introduce a variant of our method that is configured with greedy search in text generation. We denote this variant as \textbf{FastLog\textsubscript{gs}}, and denote our original method that uses beam search in text generation as \textbf{FastLog\textsubscript{bs}}.

\subsection{Evaluation Metrics}
We follow previous work \cite{what_mastropaolo2022using} to adopt \textbf{Accuracy} as the basic evaluation metric for logging positions, log levels, and log messages. When using Accuracy to evaluate predicted logging positions and log levels, Accuracy is calculated as the ratio of correctly predicted samples for logging positions or log levels. When using Accuracy to evaluate the generated texts (i.e., log messages), Accuracy refers to the proportion of samples that are exactly the same as the target log message.

Besides, we also use the text similarity metrics \textbf{BLEU} \cite{bleu_papineni2002bleu} and \textbf{ROUGE} \cite{rouge_lin2004rouge} to evaluate the quality of the generated log messages, as done in previous works \cite{what_he2018characterizing, what_ding2022logentext}.

\section{Results and Analysis}\label{sec_results}

In this section, we discuss the results of evaluating FastLog for generating and inserting complete logging statements by answering four research questions.

\begin{table}[]
\centering
\caption{Prediction Accuracy Comparison Among Different Methods in Generating Logging Statements}
\label{prediction_accuracy}
\setlength\tabcolsep{2.3pt}
\begin{tabular}{llcccc}
\toprule
\textbf{Test Set}              & \textbf{Methods}                & \textbf{Position}         & \textbf{Level}            & \textbf{Message}          & \textbf{All\,3\,Aspects}    \\ \midrule
\multirow{4}{*}{Original} & LANCE\textsubscript{T5} & 65.40\%          & 66.24\%          & 16.90\%          & 15.20\%          \\
                          & LANCE\textsubscript{PL}     & 58.04\%          & 59.37\%          & 15.27\%          & 13.89\%          \\
                          & FastLog\textsubscript{gs}    & \textbf{70.52\%} & 71.40\%          & 18.20\%          & 16.47\%          \\
                          & FastLog\textsubscript{bs}   & \textbf{70.52\%} & \textbf{71.60\%} & \textbf{18.44\%} & \textbf{16.61\%} \\ \midrule
\multirow{4}{*}{New}      & LANCE\textsubscript{T5} & 52.05\%          & 54.93\%          & 2.28\%           & 1.51\%           \\
                          & LANCE\textsubscript{PL}     & 48.69\%          & 49.41\%          & 3.62\%           & 2.49\%           \\
                          & FastLog\textsubscript{gs}    & \textbf{58.84\%} & \textbf{59.75\%} & 4.40\%           & \textbf{3.11\%}  \\
                          & FastLog\textsubscript{bs}  & \textbf{58.84\%} & 59.63\%          & \textbf{4.52\%}  & \textbf{3.11\%}  \\ \bottomrule
\end{tabular}
\end{table}

\begin{table*}[]
\centering
\caption{Performance Comparison among Different Methods in Generating Log Messages}
\label{performance_messages}
\begin{tabular}{llcccccccc}
\toprule
\textbf{Test Set}               & \textbf{Methods}                & \textbf{BLEU}           & \textbf{BLEU-1}         & \textbf{BLEU-2}         & \textbf{BLEU-3}         & \textbf{BLEU-4}         & \textbf{ROUGE-1}        & \textbf{ROUGE-2}        & \textbf{ROUGE-L}        \\ \midrule
{}                           & {LANCE\textsubscript{T5}} & {30.05}          & {48.18}          & {31.63}          & {25.50}          & {20.98}          & {54.92}          & {34.97}          & {54.62}          \\
{}                           & {LANCE\textsubscript{PL}}     & {26.13}          & {42.43}          & {27.75}          & {22.09}          & {17.93}          & {49.37}          & {31.73}          & {49.01}          \\
{}                           & {FastLog\textsubscript{gs}}    & {32.58}          & {52.02}          & {34.41}          & {27.69}          & {22.74}          & {58.70}          & {37.88}          & {58.28}          \\
\multirow{-4}{*}{{Original}} & {FastLog\textsubscript{bs}}  & {\textbf{33.43}} & {\textbf{52.40}} & {\textbf{35.21}} & {\textbf{28.58}} & {\textbf{23.69}} & {\textbf{59.91}} & {\textbf{39.03}} & {\textbf{59.53}} \\ \midrule
{}                           & {LANCE\textsubscript{T5}} & {17.20}          & {37.71}          & {19.06}          & {13.04}          & {9.34}           & {44.03}          & {19.56}          & {43.61}          \\
{}                           & {LANCE\textsubscript{PL}}     & {17.68}          & {35.38}          & {19.41}          & {13.86}          & {10.26}          & {40.72}          & {20.02}          & {40.30}          \\
{}                           & {FastLog\textsubscript{gs}}    & {20.17}          & {40.99}          & {22.25}          & {15.71}          & {11.54}          & {47.69}          & {23.18}          & {47.22}          \\
\multirow{-4}{*}{{New}}      & {FastLog\textsubscript{bs}}  & {\textbf{20.90}} & {\textbf{41.30}} & {\textbf{22.96}} & {\textbf{16.43}} & {\textbf{12.23}} & {\textbf{49.02}} & {\textbf{24.30}} & {\textbf{48.57}} \\ \bottomrule
\end{tabular}
\end{table*}

\noindent \textbf{RQ1: How does FastLog compare to the state-of-the-art approach in terms of output quality?}

In this RQ, we conduct a comparison regarding the quality of logging statements generated by different methods. Specifically, we compare the quality of logging locations and logging statement contents (including log levels and log messages).

Firstly, Table~\ref{prediction_accuracy} shows the prediction accuracy of different methods. We found that FastLog\textsubscript{gs} outperforms LANCE\textsubscript{T5} and LANCE\textsubscript{PL}
on both datasets, in terms of logging position, level, message, and ``All 3 Aspects''. ``All 3 Aspects'' means that the generated logging statement has the correct logging position, log level, and log message, and thus can be directly used by developers. It can be found that FastLog\textsubscript{bs} shows an approximate 5\% accuracy improvement in logging position and log level compared to LANCE\textsubscript{T5}, along with an about 1.5\% accuracy improvement in log messages and ``All 3 Aspects''. As compared to LANCE\textsubscript{PL}, the improvement is more obvious.

Secondly, in terms of log message similarity metrics, i.e. BLEU and ROUGE, we can also see improvement from Table~\ref{performance_messages}. 
It can be found that FastLog\textsubscript{bs} increases the BLEU values by over 3 and around 5 in ROUGE-1 and ROUGE-L compared to LANCE\textsubscript{T5}, which is an effective improvement\cite{what_mastropaolo2022using, what_ding2022logentext}. Similarly, the advantage of FastLog\textsubscript{bs} over LANCE\textsubscript{PL} is more obvious.

When considering the difference between FastLog\textsubscript{bs} and FastLog\textsubscript{gs}, it can be found that the beam search decoding strategy shows a slight advantage over the greedy search strategy.
These findings collectively demonstrate that our proposed method can generate higher-quality logging statements.

As a reminder, all methods perform worse on the new dataset than on the original dataset. By deeply inspecting these two datasets, we found a possible reason that the program methods and target log messages are much longer in the new dataset than those in the original dataset. The potential different styles of the repositories and longer target text make it much harder for models to predict logging locations and generate logging statements. However, it is worth noting that our method still outperforms the baselines.

\begin{tcolorbox}
\textit{Conclusion for RQ1:} FastLog outperforms the state-of-the-art method, LANCE, in terms of output quality for generating and inserting complete logging statements on both the original and new datasets.
\end{tcolorbox}

\noindent \textbf{RQ2: How efficient is FastLog compared to the state-of-the-art approach?}

\begin{table}[]
\centering
\setlength{\tabcolsep}{11pt}
\caption{Efficiency Comparison among Different Methods in Generating Logging Statements}
\label{efficiency}
\begin{tabular}{@{}llccc@{}}
\toprule
\textbf{Test Set}     & \textbf{Methods}       & \textbf{Total} & \textbf{Stage-1} & \textbf{Stage-2} \\ \midrule
\multirow{4}{*}{Original} & LANCE\textsubscript{T5} & 2.80s          & -                & -                \\
                          & LANCE\textsubscript{PL}     & 1.28s          & -                & -                \\
                          & FastLog\textsubscript{gs}    & 0.13s          & 0.02s            & 0.11s            \\
                          & FastLog\textsubscript{bs}  & 0.22s          & 0.02s            & 0.20s            \\ \midrule
\multirow{4}{*}{New}      & LANCE\textsubscript{T5} & 2.79s          & -                & -                \\
                          & LANCE\textsubscript{PL}     & 1.38s          & -                & -                \\
                          & FastLog\textsubscript{gs}    & 0.17s          & 0.02s            & 0.15s            \\
                          & FastLog\textsubscript{bs}  & 0.24s          & 0.02s            & 0.22s            \\ \bottomrule
\end{tabular}
\begin{threeparttable}
\begin{tablenotes}
            \small
            \item[*] ``s'' refers to seconds.
        \end{tablenotes}
\end{threeparttable}
\end{table}

In RQ2, we investigate the efficiency of different methods for generating and inserting logging statements. 
We conducted the experiments on an NVIDIA GeForce RTX 3090 with an inference batch size of 1. We adopt this batch size to simulate the real-world scenarios of providing developers with instant recommendations during their coding activity. 

Table~\ref{efficiency} presents the average time that one method takes to generate the logging statement for each example in the test set. We found that our method \textbf{significantly outperforms LANCE and its variant} in terms of the efficiency, with both datasets showing similar results.

Specifically, as compared to the original LANCE\textsubscript{T5}, FastLog\textsubscript{bs} shows about 12 times faster in both the original and the new datasets. Even LANCE\textsubscript{PL} which cuts down the time cost of its original version by half, is still about 6 times slower than FastLog\textsubscript{bs}. As for the comparison between FastLog\textsubscript{gs} that adopts the same base model and decoding strategy as LANCE\textsubscript{PL}, our design also shows a significant advantage. These results clearly verify our previous argument that by decomposing the logging statement generation task into two stages, where the model only generates shorter logging statement content instead of the entire program method, the efficiency can be significantly improved.

It is obvious from the above results that \textbf{FastLog is more practical than LANCE and can even be adapted to real-time intelligent development environments that always require instant feedback}.

As a reminder, FastLog\textsubscript{bs} is slightly slower than FastLog\textsubscript{gs}, but the difference is somewhat marginal in practice. Considering the quality of the generated statements, we still recommend FastLog\textsubscript{bs} as the first choice.

\begin{tcolorbox}
\textit{Conclusion for RQ2:} FastLog is more efficient than the state-of-the-art approach LANCE in generating and inserting complete logging statements on both datasets.
\end{tcolorbox}

\noindent \textbf{RQ3: Beyond the overall accuracy, how well does FastLog predict logging positions and levels from a more granular perspective?}

Unlike BLEU or ROUGE that uses a real number to reflect different degrees of quality of log messages, the accuracy of predicting logging positions and levels records a Boolean value of each output, i.e. whether the predicted position or level is the same as the ground-truth label.
However, we argue that not all ``wrong predictions'' have the same degree of ``incorrectness''. For example, when considering log level prediction, if the ground truth is ``Fatal'', then predictions of ``Error'' and ``Info'' actually show different degrees of incorrectness and lead to different effects: In comparison to ``Error'' that shows closer meaning to the ground truth, ``Info'' is a worse prediction as it fails to convey the problem in the log content. Similarly, in the prediction of logging positions, a wrong position that is closer to the target than another wrong position works a bit better.

Therefore, in RQ3, we try to analyze the effectiveness of different methods in predicting logging positions and levels beyond the overall accuracy. This includes a comprehensive evaluation of the distance between the prediction and target values.

In our investigation, we use the following distance metrics to show the degrees of incorrectness.
First, log levels can be ranked based on their severity as (1) Trace, (2) Debug, (3) Info, (4) Warn, (5) Error, (6) Fatal. We calculate the distance as the absolute difference between predicted and target values based on their ordinal rankings. For log positions, we calculate the distance as the absolute difference between the indices of the predicted and target tokens. When the calculated distance is zero, it indicates a correct prediction.

\begin{figure}[htbp]
\centerline{\includegraphics[width=0.49\textwidth]{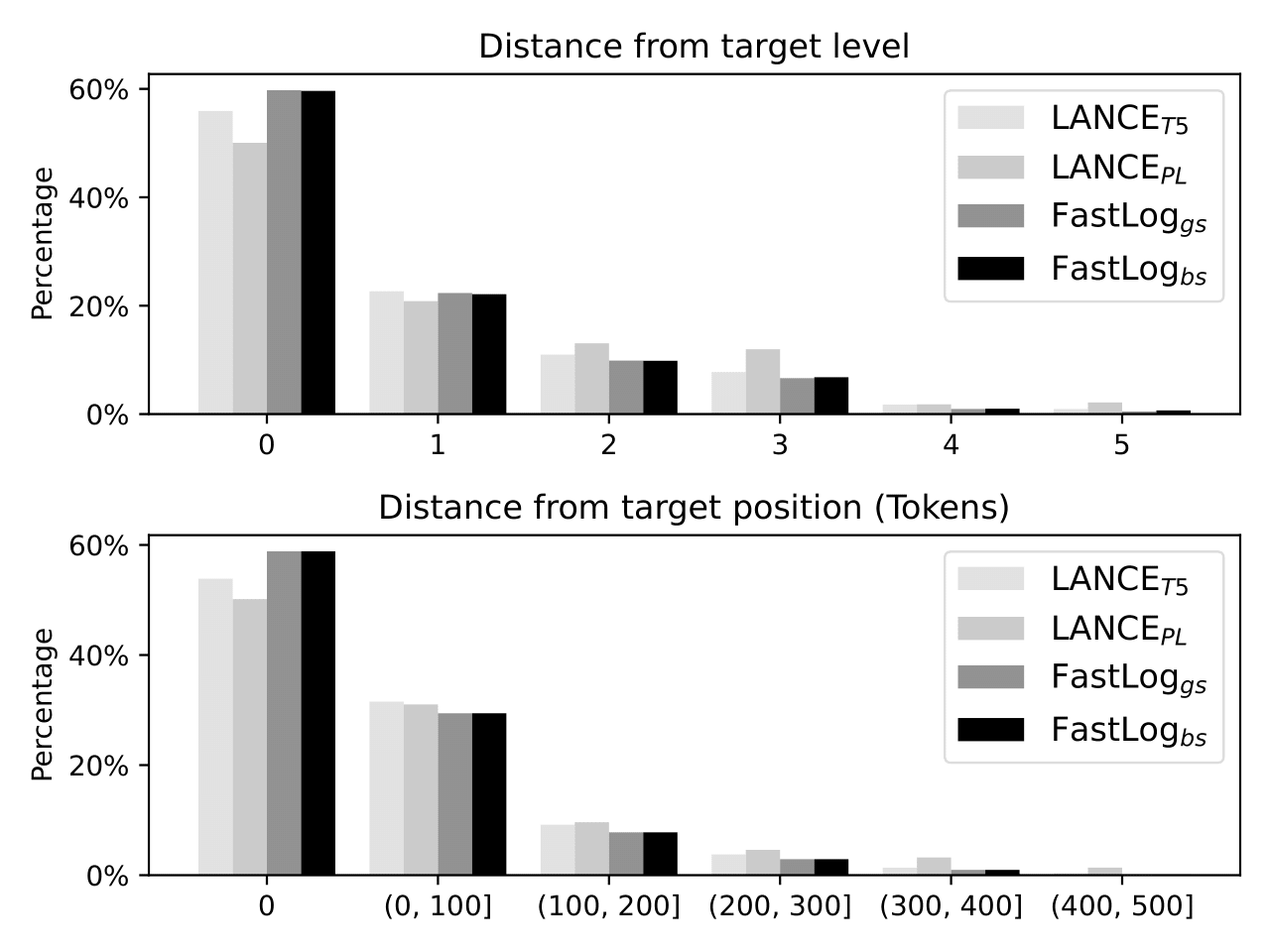}}
\caption{Distance Between Prediction and Target Values on Logging Position and Level (``0'' Indicates the Prediction is Correct.)}
\label{difference}
\end{figure}

Fig.~\ref{difference} illustrates the distance between prediction and target values for log level suggestion using different methods. (As a reminder, in RQ3 and RQ4, conclusions drawn from the original test dataset and the new test dataset are consistent with each other. Due to the space limit, we only present results on the new test dataset. For the complete results, please refer to our online artifact \cite{DATARELEASE}.)

We found that a significant portion of the predictions by FastLog\textsubscript{bs} exhibit a small distance ($\leq$1) to the target values, with 9.83\% (711) predictions having a distance of 2 and 8.43\% (610) predictions exceeding 2. However, for LANCE, especially for LANCE\textsubscript{PL}, there are more predictions with a distance over 1 to the target values, with 13.05\% (944) predictions having a distance of 2 and 15.89\% (1150) predictions exceeding 2. This suggests that our method performs better in predicting log levels since it can reflect the severity of log content more effectively.

Fig.~\ref{difference} also presents the distribution of distances between predicted and target positions at the token level.
We noticed that our method produces fewer predictions whose distance to the target value is relatively large. Compared to LANCE\textsubscript{PL}, which has 18.81\% (1361) predictions showing a distance over 100 tokens from the target position, FastLog\textsubscript{bs} only has 11.75\% (850), which suggests better quality of our method in predicting logging positions.

\begin{tcolorbox}
\textit{Conclusion for RQ3:} The prediction results of FastLog for logging positions and levels are closer to the target values, which further verifies the advantages of our method.
\end{tcolorbox}

\begin{table*}[]
\centering
\setlength{\tabcolsep}{3pt}
\begin{threeparttable}
\caption{Performance Comparison of Different Splitting Strategies in Logging Location Prediction}
\label{split}
\begin{tabular}{@{}lcccccc@{}}
\toprule
\textbf{Splitting Strategy}              & \textbf{\begin{tabular}[c]{@{}c@{}}$\boldsymbol{\leq}$512 \\ Correct Predictions\end{tabular}} & \textbf{\begin{tabular}[c]{@{}c@{}}$\boldsymbol{\leq}$512 \\ Accuracy\end{tabular}} & \textbf{\begin{tabular}[c]{@{}c@{}}\textgreater{}512 \\ Correct Predictions\end{tabular}} & \textbf{\begin{tabular}[c]{@{}c@{}}\textgreater{}512 \\ Accuracy\end{tabular}} & \textbf{\begin{tabular}[c]{@{}c@{}}Total \\ Correct Predictions\end{tabular}} & \textbf{\begin{tabular}[c]{@{}c@{}}Total \\ Accuracy\end{tabular}} \\ \midrule
truncate-discard                & \multirow{8}{*}{\textbf{3963}}                                                          & \multirow{8}{*}{\textbf{61.26\%}}                                            & 240                                                                                       & 31.33\%                                                                         & 4203                                                                          & 58.09\%                                                            \\
truncated-split               &                                                                                         &                                                                              & 183                                                                                       & 23.89\%                                                                        & 4146                                                                          & 57.30\%                                                            \\
average-split-512             &                                                                                         &                                                                              & 260                                                                                       & 33.94\%                                                                        & 4223                                                                          & 58.37\%                                                            \\
average-split-300             &                                                                                         &                                                                              & 251                                                                                       & 32.77\%                                                                        & 4214                                                                          & 58.24\%                                                            \\
average-split-300-statement-1  &                                                                                         &                                                                              & 257                                                                                       & 33.55\%                                                                        & 4220                                                                          & 58.33\%                                                            \\
average-split-300-statement-5  &                                                                                         &                                                                              & \textbf{294}                                                                              & \textbf{38.38\%}                                                               & \textbf{4257}                                                                 & \textbf{58.84\%}                                                   \\
average-split-300-statement-10 &                                                                                         &                                                                              & 292                                                                                       & 38.12\%                                                                        & 4255                                                                          & 58.81\%                                                            \\ \bottomrule
\end{tabular}
\begin{tablenotes}
            \scriptsize

            \item[*] ``$\leq$512'' means that the token length of the input is no more than 512, and ``\textgreater{}512'' indicates that it is greater than 512.
            
            \item[*] The numbers of samples in the ``$\leq$512'', ``\textgreater{}512'', and ``Total'' categories are 6469, 766, and 7235, respectively.
        \end{tablenotes}
\end{threeparttable}
\end{table*}

\begin{table}[]
\centering
\setlength{\tabcolsep}{2.5pt}
\caption{Effect of Splitting Long Input Text on Generating Logging Statements}
\label{split_logging_content}
\begin{tabular}{@{}lccc@{}}
\toprule
\textbf{Splitting Strategy}             & \textbf{Level} & \textbf{Message} & \textbf{All\,3\,Aspects} \\ \midrule
truncate-discard               & 59.36\%        & 4.46\%           & 3.04\%                 \\
average-split-300-statement-5 & \textbf{59.63\%}        & \textbf{4.52\%}           & \textbf{3.11\%}                 \\ \bottomrule
\end{tabular}
\end{table}

\noindent \textbf{RQ4: Does splitting long input text into smaller chunks benefit the generation and insertion of logging statements?}

In this RQ, we explore the effect of different splitting strategies and find out whether the prediction accuracy of insertion positions and the quality of generated logging statements can be improved by splitting long input texts.

As mentioned in Section~\ref{sec_methodology}, for the input texts whose length exceed \textit{$\text{max\_model\_input\_length}$}, we split them into smaller chunks to obtain complete prediction results of all tokens in the input. In our experiments, we set \textit{$\text{max\_model\_input\_length}$} to 512 following previous work \cite{what_mastropaolo2022using}. 

We compare a few splitting strategies: (1) ``\textbf{truncate-discard}'', which is the most naive strategy in practice, truncates and discards the tokens exceeding \textit{$\text{max\_model\_input\_length}$}; (2) ``\textbf{truncate-split}'' truncates and splits the input using \textit{$\text{max\_model\_input\_length}$} as the step size; (3) ``\textbf{average-split-\textit{m}}'' evenly splits the input text into several chunks, with each not exceeding \textit{m} tokens; and (4) ``\textbf{average-split-\textit{m}-statement-\textit{k}}'' evenly splits the input text into chunks of no more than \textit{m} tokens, with the addition of \textit{k} complete statements to the truncated boundaries for each chunk. 
The values of \textit{m} are set in this way: First, as the \textit{$\text{max\_model\_input\_length}$} is 512, we have one option of \textit{m} as this maximum value of 512. Secondly, to include additional contextual statements within this maximum input length and avoid generating too many chunks, while also ensuring that the contextual tokens do not surpass the tokens to be predicted, we set another option of \textit{m} as 300. As for \textit{k}, we set three options as 1, 5, and 10, to investigate a proper number of contextual statements.

Table~\ref{split} presents the effect of different splitting strategies on the prediction accuracy of insertion positions with FastLog\textsubscript{bs} on the new dataset\footnote{As explained in RQ3, for complete results, please refer to our online artifact \cite{DATARELEASE}.}, which shows the benefits of properly splitting long input text and adding contextual statements. The table presents the performance of various splitting strategies on the samples of different input lengths. Since text splitting is only applied to samples with a token length exceeding 512, we can focus on the accuracy of this subset of samples.

Firstly, we found that properly splitting the input text helps predict logging positions. Specifically, from Table~\ref{split}, we noticed that the ``truncate-split'' method is even worse than the most naive method ``truncate-discard''. This is because the last chunk generated by it can be very short, which severely disrupts the semantic coherence and impacts the prediction for insertion positions. Observing the last five rows in Table~\ref{split}, all of which employ an average splitting strategy, it is obvious that they all perform better than the ``truncate-discard'' and ``truncate-split'' methods. This suggests that we should averagely split the input text.

Secondly, we found that adding contextual statements is also useful. In Table~\ref{split}, the last five rows indicate that adding a proper number of contextual statements improves the prediction of where to insert logging statements. Specifically, the ``average-split-300-statement-5'' configuration increases the accuracy to 38.38\%  for the ``$>$512'' samples, outperforming the ``average-split-300'' and ``average-split-512'' methods.

Besides, we have also explored the effect of splitting long input text on generating logging statements. As the ``average-split-300-statement-5'' configuration performs best for predicting logging position, and to be consistent with Stage-1, we also use this configuration for Stage-2 to generate logging statements. From Table~\ref{split_logging_content}, which shows the effect of text splitting with FastLog\textsubscript{bs} on the new dataset, we can see that splitting long text has little effect on the logging statement content generation.

Finally, given the superior performance of the ``average-split-300-statement-5'' configuration, \textbf{we adopted this configuration for our final method, and all experiment results of the research questions above are based on it.}

\begin{tcolorbox}
\textit{Conclusion for RQ4:} Properly splitting long input text and adding contextual statements is advantageous for predicting where to insert logging statements.
\end{tcolorbox}

\section{Discussion}\label{sec_discussion}
\subsection{Modifications To Non-Log Contents in LANCE}
\label{subsec_modificationcases}
\begin{figure*}[ht]
\centerline{\includegraphics[width=1.0\textwidth]{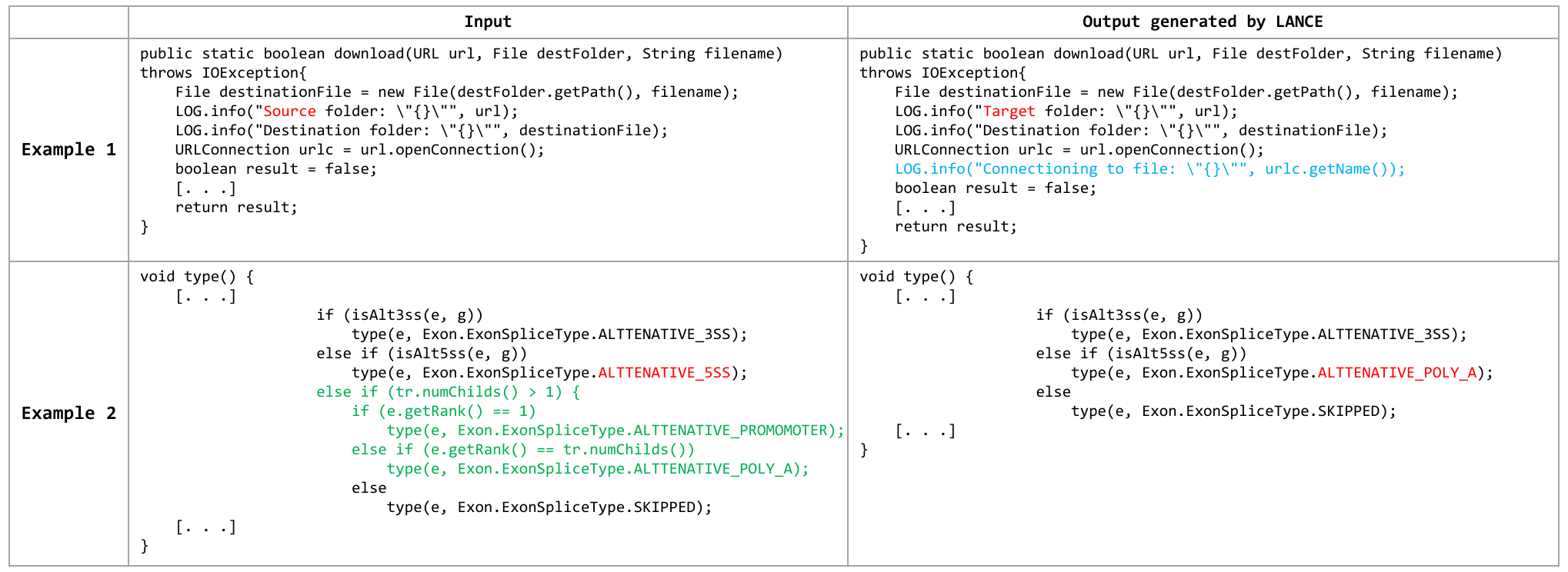}}
\caption{Examples of LANCE Modifying Non-Log Content}
\label{modify_content}
\end{figure*}
As mentioned in Section~\ref{sec_motivation}, LANCE 
needs to regenerate the entire program method, which may inadvertently result in modifications to the content other than the logging statement. Our manual analysis has identified two primary categories of such instances where LANCE makes alterations to the content other than the inserted logging statements.

The first category involves instances where LANCE inadvertently alters specific words. As depicted in Fig.~\ref{modify_content}, Example 1 serves as an illustration. In the output of LANCE, apart from the newly generated logging statement (highlighted in blue), the model also replaces the word ``source'' with ``target'' in another statement (highlighted in red).

The second category represents a more critical scenario: 
LANCE may unintentionally delete portions of the content. As exemplified by Example 2 in Fig.~\ref{modify_content}, besides changing ``ALTTENATIVE\_5SS'' to ``ALTTENATIVE\_POLY\_A'', the output of LANCE also eliminates the content highlighted in green, thereby diminishing the reliability of the output content.

In our method, since we locate the logging position through a token classification model and only focus on generating the needed logging statement, there is no need to regenerate the entire program method, \textbf{we can completely avoid modifying the content other than the logging statement}. This helps to ensure the reliability of our model.

\subsection{Enhancement by Providing Multiple Prediction Results}

\begin{table}[]
\centering
\caption{Effect of Providing Multiple Prediction Results}
\setlength{\tabcolsep}{1.3pt}
\label{multiple_accuracy}
\begin{tabular}{@{}lcccc@{}}
\toprule
\textbf{Method}              & \textbf{Position} & \textbf{Level}   & \textbf{Message} & \textbf{All\,3\,Aspects} \\ \midrule
single prediction result     & 58.84\%           & 59.63\%          & 4.52\%           & 3.11\%                 \\
10 prediction results & \textbf{83.69\%}           & \textbf{76.19\%}          & \textbf{7.89\%}  & \textbf{5.96\%}        \\ \bottomrule
\end{tabular}
\end{table}

Providing multiple prediction results has been shown as an efficient manner to increase the chance of providing the correct result for interactive tools 
\cite{multi_koyuncu2019ifixr, multi_he2021pyart, multi_cai2023properly}. 
Owing to the good efficiency of FastLog, it can be applied in Just-In-Time scenarios, and hence is eligible to support an interactive working manner. Hence, the above idea of providing multiple candidate results can be helpful in further improving the effectiveness of FastLog.
However, our method is with two stages involving two models, and each model can offer multiple prediction results. As such, we need to elaborate a proper design to combine the multiple prediction results from two stages.

We explored an approach based on the prediction probability of logging positions. When we obtain multiple predicted insertion positions from Stage-1, each position has an associated probability (i.e., the probability that a token is classified as ``1'' according to Section~\ref{sec_methodology}). By setting a probability threshold (e.g., 0.01), we retain only those predicted positions with probabilities no less than this threshold. For these collected positions, we allocate the number of generated logging statements based on their probabilities. Positions with higher probabilities can output more logging statements, and eventually, we will get a total of 10 prediction results.

The allocation process works as follows: if there are \textit{p} positions, we sort them in descending order by their probabilities. In the first pass, we allocate one logging statement to each of the first \textit{p} positions. In the second pass, we allocate one logging statement to each of the first \textit{p-1} positions, and so on.

A preliminary experiment (in Table~\ref{multiple_accuracy}) shows promising results for providing multiple predictions in generating and inserting logging statements, especially for suggesting logging positions and levels. The prediction accuracy goes up to 83.69\% and 76.19\%, respectively. \textbf{This result further demonstrates the great potential and practicality of FastLog in a real-time intelligent development environment.} After all, any  Just-In-Time scenario shall require an AI coding assistant to always give complete, high-quality, and more importantly, instant feedback.

\section{Threats to validity}\label{sec_threats}
There are three main threats to the validity of our evaluation.

\textbf{Validity of the Measurement (Construct Validity):} We use BLEU and ROUGE to assess the quality of the generated log messages. 
Although text similarity metrics may not fully capture the quality of the generated text \cite{bleu_gros2020code, bleu_roy2021reassessing}, we follow previous works \cite{what_he2018characterizing, what_ding2022logentext} to use BLEU and ROUGE as widely-accepted quality measures for the generated log messages. 

\textbf{Implementation of the Baseline (Internal Validity):} 
The second threat to validity comes from the reproducing process of the baseline. To minimize any possible inconsistency, we directly use the released model checkpoints and codes from previous work \cite{what_mastropaolo2022using}. Besides, to ensure fairness in performance comparison, we adopt the same hardware for running both our method and the baseline.

\textbf{Generalization of the Conclusions (External Validity):}
In this work, we only evaluate our method with the Java projects that use the Log4j library for logging. But since Java and Log4j are very popular programming language and logging framework, we reckon that the effectiveness of our method verified in this paper is meaningful. Besides, our methodology is not restricted to these subjects. We will explore its effects in other programming languages and logging libraries in the future.

\section{Related works}\label{sec_relatedworks}

In this section, we review the existing works on automated logging statement generation.

Some studies try to automatically determine ``where to log'' within the code snippets. Most studies treat ``where to log'' as a binary classification task at different granularities \cite{where_zhu2015learning, where_jia2018smartlog, where_candido2021exploratory, where_zhang2023deeplog}. 
Li et al. \cite{where_li2018studying} use topic modeling to find the proper methods to insert the logging statements. Li et al. \cite{where_li2020shall} design a deep learning model to suggest proper logging locations at the code-block level. 
To enhance system performance monitoring, Yao et al. \cite{where_statistic_yao2018log4perf} insert logging statements into the beginning of each method and then utilize statistical performance models to identify the locations that affect the system performance. Compared with these works, our method not only predicts the logging locations but also generates the logging statements. Besides, the first stage of our method
locates the logging positions at the finest-grained token level.

Some works focus on deciding ``what to log'', including selecting log levels and generating log messages \cite{diagnosis_yuan2012improving, what_ding2023logentext}. 
Yuan et al. \cite{what_yuan2012characterizing} check for inconsistent log levels in similar code snippets, while Mizouchi et al. \cite{what_mizouchi2019padla} dynamically adjust log levels to record detailed information when recognizing a performance anomaly. 
More directly, Li et al. \cite{what_li2021deeplv} design a neural network to suggest log levels. Later, Liu et al. \cite{what_liu2022tell} further improve the model performance by incorporating multiple levels of code block information.
Besides log levels, some works tried to generate the log text. He et al. \cite{what_he2018characterizing} design an information retrieval method, while Ding et al. \cite{what_ding2022logentext} build a neural machine translation model to generate the textual descriptions of logging statements.
For the dynamic information, Liu et al. \cite{what_liu2019variables} leverage a multi-layer neural network to learn the representation of program tokens and recommend the variables that should be logged. Compared with these works, the second stage of our method generates the complete logging statement at once, to ensure content compatibility and generation efficiency.

Recently, Mastropaolo et al. \cite{what_mastropaolo2022using} propose LANCE, a method that concurrently determines ``where to log'' and ``what to log'' by generating and inserting 
a logging statement in a given method.
LANCE is the first end-to-end method to generate and insert complete logging statements. 
Compared with LANCE, our method can accomplish this task more efficiently and effectively and thus is more practical for real-world usage.

\section{Conclusion and future work}\label{sec_conclusion}

In this paper, we propose FastLog, a two-stage method for efficiently generating and inserting complete logging statements.
Specifically, FastLog first utilizes a token classification model to locate a precise proper position to insert the logging statement and next uses a Seq2Seq model to generate a complete logging statement for that position.
By properly organizing the sub-tasks of deciding ``where to log'' and ``what to log'', FastLog significantly reduces the unnecessary time cost in text generation and avoids the risk of modifying code content other than logging statements. The model burden also tends to be reduced, which may contribute to better performance. 
Extensive experimental results demonstrate that FastLog has outstanding efficiency and higher output quality compared to the state-of-the-art methods, suggesting that FastLog is more effective in helping developers write logging statements.

In the future, we will first evaluate the effectiveness of our method in other programming languages and using different logging libraries. 
In addition, our preliminary results suggest that providing proper multiple prediction results can effectively increase the chance of our efficient method offering high-quality logging statement suggestions. Thus, we will further explore effective strategies to combine the predictions in each stage to make our method effectively help its users in interactive Just-In-Time scenarios.

\section*{Data Availability}\label{sec_data}
The artifact of this paper is online available at \cite{DATARELEASE}.

\section*{Acknowledgments}
This work was supported by National Natural Science Foundation of China (Grant No. 62250610224).


\bibliographystyle{ACM-Reference-Format}
\balance
\bibliography{reference}


\begin{thebibliography}{49}


\ifx \showCODEN    \undefined \def \showCODEN     #1{\unskip}     \fi
\ifx \showDOI      \undefined \def \showDOI       #1{#1}\fi
\ifx \showISBNx    \undefined \def \showISBNx     #1{\unskip}     \fi
\ifx \showISBNxiii \undefined \def \showISBNxiii  #1{\unskip}     \fi
\ifx \showISSN     \undefined \def \showISSN      #1{\unskip}     \fi
\ifx \showLCCN     \undefined \def \showLCCN      #1{\unskip}     \fi
\ifx \shownote     \undefined \def \shownote      #1{#1}          \fi
\ifx \showarticletitle \undefined \def \showarticletitle #1{#1}   \fi
\ifx \showURL      \undefined \def \showURL       {\relax}        \fi
\providecommand\bibfield[2]{#2}
\providecommand\bibinfo[2]{#2}
\providecommand\natexlab[1]{#1}
\providecommand\showeprint[2][]{arXiv:#2}

\bibitem[DAT({[n.\,d.]})]%
        {DATARELEASE}
 \bibinfo{year}{[n.\,d.]}\natexlab{}.
\newblock \bibinfo{title}{The replication package for this paper}.
\newblock
\newblock
\urldef\tempurl%
\url{https://github.com/zhipeng-cai/FastLog}
\showURL{%
\tempurl}


\bibitem[Ahmad et~al\mbox{.}(2021)]%
        {plbart_ahmad2021unified}
\bibfield{author}{\bibinfo{person}{Wasi~Uddin Ahmad}, \bibinfo{person}{Saikat Chakraborty}, \bibinfo{person}{Baishakhi Ray}, {and} \bibinfo{person}{Kai-Wei Chang}.} \bibinfo{year}{2021}\natexlab{}.
\newblock \showarticletitle{Unified pre-training for program understanding and generation}.
\newblock \bibinfo{journal}{\emph{arXiv preprint arXiv:2103.06333}} (\bibinfo{year}{2021}).
\newblock


\bibitem[Cai et~al\mbox{.}(2023)]%
        {multi_cai2023properly}
\bibfield{author}{\bibinfo{person}{Zhipeng Cai}, \bibinfo{person}{Songqiang Chen}, {and} \bibinfo{person}{Xiaoyuan Xie}.} \bibinfo{year}{2023}\natexlab{}.
\newblock \showarticletitle{Properly Offer Options to Improve the Practicality of Software Document Completion Tools}. In \bibinfo{booktitle}{\emph{2023 IEEE/ACM 31st International Conference on Program Comprehension (ICPC)}}. IEEE, \bibinfo{pages}{237--241}.
\newblock


\bibitem[C{\^a}ndido et~al\mbox{.}(2021)]%
        {where_candido2021exploratory}
\bibfield{author}{\bibinfo{person}{Jeanderson C{\^a}ndido}, \bibinfo{person}{Jan Haesen}, \bibinfo{person}{Maur{\'\i}cio Aniche}, {and} \bibinfo{person}{Arie van Deursen}.} \bibinfo{year}{2021}\natexlab{}.
\newblock \showarticletitle{An exploratory study of log placement recommendation in an enterprise system}. In \bibinfo{booktitle}{\emph{2021 IEEE/ACM 18th International Conference on Mining Software Repositories (MSR)}}. IEEE, \bibinfo{pages}{143--154}.
\newblock


\bibitem[Chen et~al\mbox{.}(2018)]%
        {testing_chen2018automated}
\bibfield{author}{\bibinfo{person}{Boyuan Chen}, \bibinfo{person}{Jian Song}, \bibinfo{person}{Peng Xu}, \bibinfo{person}{Xing Hu}, {and} \bibinfo{person}{Zhen~Ming Jiang}.} \bibinfo{year}{2018}\natexlab{}.
\newblock \showarticletitle{An automated approach to estimating code coverage measures via execution logs}. In \bibinfo{booktitle}{\emph{Proceedings of the 33rd ACM/IEEE International Conference on Automated Software Engineering}}. \bibinfo{pages}{305--316}.
\newblock


\bibitem[Chen et~al\mbox{.}(2017)]%
        {testing_chen2017analytics}
\bibfield{author}{\bibinfo{person}{Tse-Hsun Chen}, \bibinfo{person}{Mark~D Syer}, \bibinfo{person}{Weiyi Shang}, \bibinfo{person}{Zhen~Ming Jiang}, \bibinfo{person}{Ahmed~E Hassan}, \bibinfo{person}{Mohamed Nasser}, {and} \bibinfo{person}{Parminder Flora}.} \bibinfo{year}{2017}\natexlab{}.
\newblock \showarticletitle{Analytics-driven load testing: An industrial experience report on load testing of large-scale systems}. In \bibinfo{booktitle}{\emph{2017 IEEE/ACM 39th International Conference on Software Engineering: Software Engineering in Practice Track (ICSE-SEIP)}}. IEEE, \bibinfo{pages}{243--252}.
\newblock


\bibitem[Dabic et~al\mbox{.}(2021)]%
        {dataset_dabic2021sampling}
\bibfield{author}{\bibinfo{person}{Ozren Dabic}, \bibinfo{person}{Emad Aghajani}, {and} \bibinfo{person}{Gabriele Bavota}.} \bibinfo{year}{2021}\natexlab{}.
\newblock \showarticletitle{Sampling projects in github for MSR studies}. In \bibinfo{booktitle}{\emph{2021 IEEE/ACM 18th International Conference on Mining Software Repositories (MSR)}}. IEEE, \bibinfo{pages}{560--564}.
\newblock


\bibitem[Devlin et~al\mbox{.}(2018)]%
        {bert_devlin2018bert}
\bibfield{author}{\bibinfo{person}{Jacob Devlin}, \bibinfo{person}{Ming-Wei Chang}, \bibinfo{person}{Kenton Lee}, {and} \bibinfo{person}{Kristina Toutanova}.} \bibinfo{year}{2018}\natexlab{}.
\newblock \showarticletitle{Bert: Pre-training of deep bidirectional transformers for language understanding}.
\newblock \bibinfo{journal}{\emph{arXiv preprint arXiv:1810.04805}} (\bibinfo{year}{2018}).
\newblock


\bibitem[Ding et~al\mbox{.}(2022)]%
        {what_ding2022logentext}
\bibfield{author}{\bibinfo{person}{Zishuo Ding}, \bibinfo{person}{Heng Li}, {and} \bibinfo{person}{Weiyi Shang}.} \bibinfo{year}{2022}\natexlab{}.
\newblock \showarticletitle{Logentext: Automatically generating logging texts using neural machine translation}. In \bibinfo{booktitle}{\emph{2022 IEEE International Conference on Software Analysis, Evolution and Reengineering (SANER)}}. IEEE, \bibinfo{pages}{349--360}.
\newblock


\bibitem[Ding et~al\mbox{.}(2023)]%
        {what_ding2023logentext}
\bibfield{author}{\bibinfo{person}{Zishuo Ding}, \bibinfo{person}{Yiming Tang}, \bibinfo{person}{Xiaoyu Cheng}, \bibinfo{person}{Heng Li}, {and} \bibinfo{person}{Weiyi Shang}.} \bibinfo{year}{2023}\natexlab{}.
\newblock \showarticletitle{LoGenText-Plus: Improving Neural Machine Translation-based Logging Texts Generation with Syntactic Templates}.
\newblock \bibinfo{journal}{\emph{ACM Transactions on Software Engineering and Methodology}} (\bibinfo{year}{2023}).
\newblock


\bibitem[Du et~al\mbox{.}(2017)]%
        {anomaly_du2017deeplog}
\bibfield{author}{\bibinfo{person}{Min Du}, \bibinfo{person}{Feifei Li}, \bibinfo{person}{Guineng Zheng}, {and} \bibinfo{person}{Vivek Srikumar}.} \bibinfo{year}{2017}\natexlab{}.
\newblock \showarticletitle{Deeplog: Anomaly detection and diagnosis from system logs through deep learning}. In \bibinfo{booktitle}{\emph{Proceedings of the 2017 ACM SIGSAC conference on computer and communications security}}. \bibinfo{pages}{1285--1298}.
\newblock


\bibitem[Fu et~al\mbox{.}(2014)]%
        {challenge_where_fu2014developers}
\bibfield{author}{\bibinfo{person}{Qiang Fu}, \bibinfo{person}{Jieming Zhu}, \bibinfo{person}{Wenlu Hu}, \bibinfo{person}{Jian-Guang Lou}, \bibinfo{person}{Rui Ding}, \bibinfo{person}{Qingwei Lin}, \bibinfo{person}{Dongmei Zhang}, {and} \bibinfo{person}{Tao Xie}.} \bibinfo{year}{2014}\natexlab{}.
\newblock \showarticletitle{Where do developers log? an empirical study on logging practices in industry}. In \bibinfo{booktitle}{\emph{Companion Proceedings of the 36th International Conference on Software Engineering}}. \bibinfo{pages}{24--33}.
\newblock


\bibitem[Giulianelli et~al\mbox{.}(2023)]%
        {uncertainty_giulianelli2023comes}
\bibfield{author}{\bibinfo{person}{Mario Giulianelli}, \bibinfo{person}{Joris Baan}, \bibinfo{person}{Wilker Aziz}, \bibinfo{person}{Raquel Fern{\'a}ndez}, {and} \bibinfo{person}{Barbara Plank}.} \bibinfo{year}{2023}\natexlab{}.
\newblock \showarticletitle{What Comes Next? Evaluating Uncertainty in Neural Text Generators Against Human Production Variability}.
\newblock \bibinfo{journal}{\emph{arXiv preprint arXiv:2305.11707}} (\bibinfo{year}{2023}).
\newblock


\bibitem[Gros et~al\mbox{.}(2020)]%
        {bleu_gros2020code}
\bibfield{author}{\bibinfo{person}{David Gros}, \bibinfo{person}{Hariharan Sezhiyan}, \bibinfo{person}{Prem Devanbu}, {and} \bibinfo{person}{Zhou Yu}.} \bibinfo{year}{2020}\natexlab{}.
\newblock \showarticletitle{Code to comment" translation" data, metrics, baselining \& evaluation}. In \bibinfo{booktitle}{\emph{Proceedings of the 35th IEEE/ACM International Conference on Automated Software Engineering}}. \bibinfo{pages}{746--757}.
\newblock


\bibitem[Harty et~al\mbox{.}(2021)]%
        {monitoring_harty2021logging}
\bibfield{author}{\bibinfo{person}{Julian Harty}, \bibinfo{person}{Haonan Zhang}, \bibinfo{person}{Lili Wei}, \bibinfo{person}{Luca Pascarella}, \bibinfo{person}{Mauricio Aniche}, {and} \bibinfo{person}{Weiyi Shang}.} \bibinfo{year}{2021}\natexlab{}.
\newblock \showarticletitle{Logging practices with mobile analytics: An empirical study on firebase}. In \bibinfo{booktitle}{\emph{2021 IEEE/ACM 8th International Conference on Mobile Software Engineering and Systems (MobileSoft)}}. IEEE, \bibinfo{pages}{56--60}.
\newblock


\bibitem[Hasselbring and van Hoorn(2020)]%
        {monitoring_hasselbring2020kieker}
\bibfield{author}{\bibinfo{person}{Wilhelm Hasselbring} {and} \bibinfo{person}{Andr{\'e} van Hoorn}.} \bibinfo{year}{2020}\natexlab{}.
\newblock \showarticletitle{Kieker: A monitoring framework for software engineering research}.
\newblock \bibinfo{journal}{\emph{Software Impacts}}  \bibinfo{volume}{5} (\bibinfo{year}{2020}), \bibinfo{pages}{100019}.
\newblock


\bibitem[He et~al\mbox{.}(2018)]%
        {what_he2018characterizing}
\bibfield{author}{\bibinfo{person}{Pinjia He}, \bibinfo{person}{Zhuangbin Chen}, \bibinfo{person}{Shilin He}, {and} \bibinfo{person}{Michael~R Lyu}.} \bibinfo{year}{2018}\natexlab{}.
\newblock \showarticletitle{Characterizing the natural language descriptions in software logging statements}. In \bibinfo{booktitle}{\emph{Proceedings of the 33rd ACM/IEEE International Conference on Automated Software Engineering}}. \bibinfo{pages}{178--189}.
\newblock


\bibitem[He et~al\mbox{.}(2021)]%
        {multi_he2021pyart}
\bibfield{author}{\bibinfo{person}{Xincheng He}, \bibinfo{person}{Lei Xu}, \bibinfo{person}{Xiangyu Zhang}, \bibinfo{person}{Rui Hao}, \bibinfo{person}{Yang Feng}, {and} \bibinfo{person}{Baowen Xu}.} \bibinfo{year}{2021}\natexlab{}.
\newblock \showarticletitle{Pyart: Python api recommendation in real-time}. In \bibinfo{booktitle}{\emph{2021 IEEE/ACM 43rd International Conference on Software Engineering (ICSE)}}. IEEE, \bibinfo{pages}{1634--1645}.
\newblock


\bibitem[Jia et~al\mbox{.}(2018)]%
        {where_jia2018smartlog}
\bibfield{author}{\bibinfo{person}{Zhouyang Jia}, \bibinfo{person}{Shanshan Li}, \bibinfo{person}{Xiaodong Liu}, \bibinfo{person}{Xiangke Liao}, {and} \bibinfo{person}{Yunhuai Liu}.} \bibinfo{year}{2018}\natexlab{}.
\newblock \showarticletitle{SMARTLOG: Place error log statement by deep understanding of log intention}. In \bibinfo{booktitle}{\emph{2018 IEEE 25th International Conference on Software Analysis, Evolution and Reengineering (SANER)}}. IEEE, \bibinfo{pages}{61--71}.
\newblock


\bibitem[Koyuncu et~al\mbox{.}(2019)]%
        {multi_koyuncu2019ifixr}
\bibfield{author}{\bibinfo{person}{Anil Koyuncu}, \bibinfo{person}{Kui Liu}, \bibinfo{person}{Tegawend{\'e}~F Bissyand{\'e}}, \bibinfo{person}{Dongsun Kim}, \bibinfo{person}{Martin Monperrus}, \bibinfo{person}{Jacques Klein}, {and} \bibinfo{person}{Yves Le~Traon}.} \bibinfo{year}{2019}\natexlab{}.
\newblock \showarticletitle{iFixR: Bug report driven program repair}. In \bibinfo{booktitle}{\emph{Proceedings of the 2019 27th ACM joint meeting on european software engineering conference and symposium on the foundations of software engineering}}. \bibinfo{pages}{314--325}.
\newblock


\bibitem[Lewis et~al\mbox{.}(2019)]%
        {bart_lewis2019bart}
\bibfield{author}{\bibinfo{person}{Mike Lewis}, \bibinfo{person}{Yinhan Liu}, \bibinfo{person}{Naman Goyal}, \bibinfo{person}{Marjan Ghazvininejad}, \bibinfo{person}{Abdelrahman Mohamed}, \bibinfo{person}{Omer Levy}, \bibinfo{person}{Ves Stoyanov}, {and} \bibinfo{person}{Luke Zettlemoyer}.} \bibinfo{year}{2019}\natexlab{}.
\newblock \showarticletitle{Bart: Denoising sequence-to-sequence pre-training for natural language generation, translation, and comprehension}.
\newblock \bibinfo{journal}{\emph{arXiv preprint arXiv:1910.13461}} (\bibinfo{year}{2019}).
\newblock


\bibitem[Li et~al\mbox{.}(2018)]%
        {where_li2018studying}
\bibfield{author}{\bibinfo{person}{Heng Li}, \bibinfo{person}{Tse-Hsun Chen}, \bibinfo{person}{Weiyi Shang}, {and} \bibinfo{person}{Ahmed~E Hassan}.} \bibinfo{year}{2018}\natexlab{}.
\newblock \showarticletitle{Studying software logging using topic models}.
\newblock \bibinfo{journal}{\emph{Empirical Software Engineering}}  \bibinfo{volume}{23} (\bibinfo{year}{2018}), \bibinfo{pages}{2655--2694}.
\newblock


\bibitem[Li et~al\mbox{.}(2020b)]%
        {challenge_level_li2020qualitative}
\bibfield{author}{\bibinfo{person}{Heng Li}, \bibinfo{person}{Weiyi Shang}, \bibinfo{person}{Bram Adams}, \bibinfo{person}{Mohammed Sayagh}, {and} \bibinfo{person}{Ahmed~E Hassan}.} \bibinfo{year}{2020}\natexlab{b}.
\newblock \showarticletitle{A qualitative study of the benefits and costs of logging from developers’ perspectives}.
\newblock \bibinfo{journal}{\emph{IEEE Transactions on Software Engineering}} \bibinfo{volume}{47}, \bibinfo{number}{12} (\bibinfo{year}{2020}), \bibinfo{pages}{2858--2873}.
\newblock


\bibitem[Li et~al\mbox{.}(2017)]%
        {challenge_level_li2017log}
\bibfield{author}{\bibinfo{person}{Heng Li}, \bibinfo{person}{Weiyi Shang}, {and} \bibinfo{person}{Ahmed~E Hassan}.} \bibinfo{year}{2017}\natexlab{}.
\newblock \showarticletitle{Which log level should developers choose for a new logging statement?}
\newblock \bibinfo{journal}{\emph{Empirical Software Engineering}}  \bibinfo{volume}{22} (\bibinfo{year}{2017}), \bibinfo{pages}{1684--1716}.
\newblock


\bibitem[Li et~al\mbox{.}(2020a)]%
        {where_li2020shall}
\bibfield{author}{\bibinfo{person}{Zhenhao Li}, \bibinfo{person}{Tse-Hsun Chen}, {and} \bibinfo{person}{Weiyi Shang}.} \bibinfo{year}{2020}\natexlab{a}.
\newblock \showarticletitle{Where shall we log? studying and suggesting logging locations in code blocks}. In \bibinfo{booktitle}{\emph{Proceedings of the 35th IEEE/ACM International Conference on Automated Software Engineering}}. \bibinfo{pages}{361--372}.
\newblock


\bibitem[Li et~al\mbox{.}(2021)]%
        {what_li2021deeplv}
\bibfield{author}{\bibinfo{person}{Zhenhao Li}, \bibinfo{person}{Heng Li}, \bibinfo{person}{Tse-Hsun Chen}, {and} \bibinfo{person}{Weiyi Shang}.} \bibinfo{year}{2021}\natexlab{}.
\newblock \showarticletitle{DeepLV: Suggesting log levels using ordinal based neural networks}. In \bibinfo{booktitle}{\emph{2021 IEEE/ACM 43rd International Conference on Software Engineering (ICSE)}}. IEEE, \bibinfo{pages}{1461--1472}.
\newblock


\bibitem[Liang et~al\mbox{.}(2023)]%
        {long_liang2023open}
\bibfield{author}{\bibinfo{person}{Xiaobo Liang}, \bibinfo{person}{Zecheng Tang}, \bibinfo{person}{Juntao Li}, {and} \bibinfo{person}{Min Zhang}.} \bibinfo{year}{2023}\natexlab{}.
\newblock \showarticletitle{Open-ended Long Text Generation via Masked Language Modeling}. In \bibinfo{booktitle}{\emph{Proceedings of the 61st Annual Meeting of the Association for Computational Linguistics (Volume 1: Long Papers)}}. \bibinfo{pages}{223--241}.
\newblock


\bibitem[Lin(2004)]%
        {rouge_lin2004rouge}
\bibfield{author}{\bibinfo{person}{Chin-Yew Lin}.} \bibinfo{year}{2004}\natexlab{}.
\newblock \showarticletitle{Rouge: A package for automatic evaluation of summaries}. In \bibinfo{booktitle}{\emph{Text summarization branches out}}. \bibinfo{pages}{74--81}.
\newblock


\bibitem[Liu et~al\mbox{.}(2022)]%
        {what_liu2022tell}
\bibfield{author}{\bibinfo{person}{Jiahao Liu}, \bibinfo{person}{Jun Zeng}, \bibinfo{person}{Xiang Wang}, \bibinfo{person}{Kaihang Ji}, {and} \bibinfo{person}{Zhenkai Liang}.} \bibinfo{year}{2022}\natexlab{}.
\newblock \showarticletitle{Tell: log level suggestions via modeling multi-level code block information}. In \bibinfo{booktitle}{\emph{Proceedings of the 31st ACM SIGSOFT International Symposium on Software Testing and Analysis}}. \bibinfo{pages}{27--38}.
\newblock


\bibitem[Liu et~al\mbox{.}(2019)]%
        {what_liu2019variables}
\bibfield{author}{\bibinfo{person}{Zhongxin Liu}, \bibinfo{person}{Xin Xia}, \bibinfo{person}{David Lo}, \bibinfo{person}{Zhenchang Xing}, \bibinfo{person}{Ahmed~E Hassan}, {and} \bibinfo{person}{Shanping Li}.} \bibinfo{year}{2019}\natexlab{}.
\newblock \showarticletitle{Which variables should i log?}
\newblock \bibinfo{journal}{\emph{IEEE Transactions on Software Engineering}} \bibinfo{volume}{47}, \bibinfo{number}{9} (\bibinfo{year}{2019}), \bibinfo{pages}{2012--2031}.
\newblock


\bibitem[Lu et~al\mbox{.}({[n.\,d.]})]%
        {anomaly_ludetecting}
\bibfield{author}{\bibinfo{person}{Siyang Lu}, \bibinfo{person}{Xiang Wei}, \bibinfo{person}{Yandong Li}, {and} \bibinfo{person}{Liqiang Wang}.} \bibinfo{year}{[n.\,d.]}\natexlab{}.
\newblock \showarticletitle{Detecting anomaly in big data system logs using convolutional neural network. In 2018 IEEE 16th Intl Conf on Dependable, Autonomic and Secure Computing, 16th Intl Conf on Pervasive Intelligence and Computing}. In \bibinfo{booktitle}{\emph{4th Intl Conf on Big Data Intelligence and Computing and Cyber Science and Technology Congress (DASC/PiCom/DataCom/CyberSciTech)}}. \bibinfo{pages}{151--158}.
\newblock


\bibitem[Mastropaolo et~al\mbox{.}(2022)]%
        {what_mastropaolo2022using}
\bibfield{author}{\bibinfo{person}{Antonio Mastropaolo}, \bibinfo{person}{Luca Pascarella}, {and} \bibinfo{person}{Gabriele Bavota}.} \bibinfo{year}{2022}\natexlab{}.
\newblock \showarticletitle{Using deep learning to generate complete log statements}. In \bibinfo{booktitle}{\emph{Proceedings of the 44th International Conference on Software Engineering}}. \bibinfo{pages}{2279--2290}.
\newblock


\bibitem[Meng et~al\mbox{.}(2019)]%
        {anomaly_meng2019loganomaly}
\bibfield{author}{\bibinfo{person}{Weibin Meng}, \bibinfo{person}{Ying Liu}, \bibinfo{person}{Yichen Zhu}, \bibinfo{person}{Shenglin Zhang}, \bibinfo{person}{Dan Pei}, \bibinfo{person}{Yuqing Liu}, \bibinfo{person}{Yihao Chen}, \bibinfo{person}{Ruizhi Zhang}, \bibinfo{person}{Shimin Tao}, \bibinfo{person}{Pei Sun}, {et~al\mbox{.}}} \bibinfo{year}{2019}\natexlab{}.
\newblock \showarticletitle{Loganomaly: Unsupervised detection of sequential and quantitative anomalies in unstructured logs.}. In \bibinfo{booktitle}{\emph{IJCAI}}, Vol.~\bibinfo{volume}{19}. \bibinfo{pages}{4739--4745}.
\newblock


\bibitem[Mizouchi et~al\mbox{.}(2019)]%
        {what_mizouchi2019padla}
\bibfield{author}{\bibinfo{person}{Tsuyoshi Mizouchi}, \bibinfo{person}{Kazumasa Shimari}, \bibinfo{person}{Takashi Ishio}, {and} \bibinfo{person}{Katsuro Inoue}.} \bibinfo{year}{2019}\natexlab{}.
\newblock \showarticletitle{PADLA: a dynamic log level adapter using online phase detection}. In \bibinfo{booktitle}{\emph{2019 IEEE/ACM 27th International Conference on Program Comprehension (ICPC)}}. IEEE, \bibinfo{pages}{135--138}.
\newblock


\bibitem[Papineni et~al\mbox{.}(2002)]%
        {bleu_papineni2002bleu}
\bibfield{author}{\bibinfo{person}{Kishore Papineni}, \bibinfo{person}{Salim Roukos}, \bibinfo{person}{Todd Ward}, {and} \bibinfo{person}{Wei-Jing Zhu}.} \bibinfo{year}{2002}\natexlab{}.
\newblock \showarticletitle{Bleu: a method for automatic evaluation of machine translation}. In \bibinfo{booktitle}{\emph{Proceedings of the 40th annual meeting of the Association for Computational Linguistics}}. \bibinfo{pages}{311--318}.
\newblock


\bibitem[Raffel et~al\mbox{.}(2020)]%
        {t5_raffel2020exploring}
\bibfield{author}{\bibinfo{person}{Colin Raffel}, \bibinfo{person}{Noam Shazeer}, \bibinfo{person}{Adam Roberts}, \bibinfo{person}{Katherine Lee}, \bibinfo{person}{Sharan Narang}, \bibinfo{person}{Michael Matena}, \bibinfo{person}{Yanqi Zhou}, \bibinfo{person}{Wei Li}, {and} \bibinfo{person}{Peter~J Liu}.} \bibinfo{year}{2020}\natexlab{}.
\newblock \showarticletitle{Exploring the limits of transfer learning with a unified text-to-text transformer}.
\newblock \bibinfo{journal}{\emph{The Journal of Machine Learning Research}} \bibinfo{volume}{21}, \bibinfo{number}{1} (\bibinfo{year}{2020}), \bibinfo{pages}{5485--5551}.
\newblock


\bibitem[Roy et~al\mbox{.}(2021)]%
        {bleu_roy2021reassessing}
\bibfield{author}{\bibinfo{person}{Devjeet Roy}, \bibinfo{person}{Sarah Fakhoury}, {and} \bibinfo{person}{Venera Arnaoudova}.} \bibinfo{year}{2021}\natexlab{}.
\newblock \showarticletitle{Reassessing automatic evaluation metrics for code summarization tasks}. In \bibinfo{booktitle}{\emph{Proceedings of the 29th ACM Joint Meeting on European Software Engineering Conference and Symposium on the Foundations of Software Engineering}}. \bibinfo{pages}{1105--1116}.
\newblock


\bibitem[Satyanarayanan et~al\mbox{.}(1992)]%
        {debugging_satyanarayanan1992transparent}
\bibfield{author}{\bibinfo{person}{Mahadev Satyanarayanan}, \bibinfo{person}{David~C Steere}, \bibinfo{person}{Masashi Kudo}, {and} \bibinfo{person}{Hank Mashburn}.} \bibinfo{year}{1992}\natexlab{}.
\newblock \showarticletitle{Transparent logging as a technique for debugging complex distributed systems}. In \bibinfo{booktitle}{\emph{Proceedings of the 5th workshop on ACM SIGOPS European workshop: Models and paradigms for distributed systems structuring}}. \bibinfo{pages}{1--3}.
\newblock


\bibitem[Yang et~al\mbox{.}(2021)]%
        {anomaly_yang2021semi}
\bibfield{author}{\bibinfo{person}{Lin Yang}, \bibinfo{person}{Junjie Chen}, \bibinfo{person}{Zan Wang}, \bibinfo{person}{Weijing Wang}, \bibinfo{person}{Jiajun Jiang}, \bibinfo{person}{Xuyuan Dong}, {and} \bibinfo{person}{Wenbin Zhang}.} \bibinfo{year}{2021}\natexlab{}.
\newblock \showarticletitle{Semi-supervised log-based anomaly detection via probabilistic label estimation}. In \bibinfo{booktitle}{\emph{2021 IEEE/ACM 43rd International Conference on Software Engineering (ICSE)}}. IEEE, \bibinfo{pages}{1448--1460}.
\newblock


\bibitem[Yao et~al\mbox{.}(2018)]%
        {where_statistic_yao2018log4perf}
\bibfield{author}{\bibinfo{person}{Kundi Yao}, \bibinfo{person}{Guilherme B.~de P{\'a}dua}, \bibinfo{person}{Weiyi Shang}, \bibinfo{person}{Steve Sporea}, \bibinfo{person}{Andrei Toma}, {and} \bibinfo{person}{Sarah Sajedi}.} \bibinfo{year}{2018}\natexlab{}.
\newblock \showarticletitle{Log4perf: Suggesting logging locations for web-based systems' performance monitoring}. In \bibinfo{booktitle}{\emph{Proceedings of the 2018 ACM/SPEC International Conference on Performance Engineering}}. \bibinfo{pages}{127--138}.
\newblock


\bibitem[Yuan et~al\mbox{.}(2010)]%
        {diagnosis_yuan2010sherlog}
\bibfield{author}{\bibinfo{person}{Ding Yuan}, \bibinfo{person}{Haohui Mai}, \bibinfo{person}{Weiwei Xiong}, \bibinfo{person}{Lin Tan}, \bibinfo{person}{Yuanyuan Zhou}, {and} \bibinfo{person}{Shankar Pasupathy}.} \bibinfo{year}{2010}\natexlab{}.
\newblock \showarticletitle{Sherlog: error diagnosis by connecting clues from run-time logs}. In \bibinfo{booktitle}{\emph{Proceedings of the fifteenth International Conference on Architectural support for programming languages and operating systems}}. \bibinfo{pages}{143--154}.
\newblock


\bibitem[Yuan et~al\mbox{.}(2012a)]%
        {what_yuan2012characterizing}
\bibfield{author}{\bibinfo{person}{Ding Yuan}, \bibinfo{person}{Soyeon Park}, {and} \bibinfo{person}{Yuanyuan Zhou}.} \bibinfo{year}{2012}\natexlab{a}.
\newblock \showarticletitle{Characterizing logging practices in open-source software}. In \bibinfo{booktitle}{\emph{2012 34th International Conference on Software Engineering (ICSE)}}. IEEE, \bibinfo{pages}{102--112}.
\newblock


\bibitem[Yuan et~al\mbox{.}(2012b)]%
        {diagnosis_yuan2012improving}
\bibfield{author}{\bibinfo{person}{Ding Yuan}, \bibinfo{person}{Jing Zheng}, \bibinfo{person}{Soyeon Park}, \bibinfo{person}{Yuanyuan Zhou}, {and} \bibinfo{person}{Stefan Savage}.} \bibinfo{year}{2012}\natexlab{b}.
\newblock \showarticletitle{Improving software diagnosability via log enhancement}.
\newblock \bibinfo{journal}{\emph{ACM Transactions on Computer Systems (TOCS)}} \bibinfo{volume}{30}, \bibinfo{number}{1} (\bibinfo{year}{2012}), \bibinfo{pages}{1--28}.
\newblock


\bibitem[Zeng et~al\mbox{.}(2019)]%
        {challenge_where_zeng2019studying}
\bibfield{author}{\bibinfo{person}{Yi Zeng}, \bibinfo{person}{Jinfu Chen}, \bibinfo{person}{Weiyi Shang}, {and} \bibinfo{person}{Tse-Hsun Chen}.} \bibinfo{year}{2019}\natexlab{}.
\newblock \showarticletitle{Studying the characteristics of logging practices in mobile apps: a case study on f-droid}.
\newblock \bibinfo{journal}{\emph{Empirical Software Engineering}}  \bibinfo{volume}{24} (\bibinfo{year}{2019}), \bibinfo{pages}{3394--3434}.
\newblock


\bibitem[Zhang et~al\mbox{.}(2019)]%
        {anomaly_zhang2019robust}
\bibfield{author}{\bibinfo{person}{Xu Zhang}, \bibinfo{person}{Yong Xu}, \bibinfo{person}{Qingwei Lin}, \bibinfo{person}{Bo Qiao}, \bibinfo{person}{Hongyu Zhang}, \bibinfo{person}{Yingnong Dang}, \bibinfo{person}{Chunyu Xie}, \bibinfo{person}{Xinsheng Yang}, \bibinfo{person}{Qian Cheng}, \bibinfo{person}{Ze Li}, {et~al\mbox{.}}} \bibinfo{year}{2019}\natexlab{}.
\newblock \showarticletitle{Robust log-based anomaly detection on unstable log data}. In \bibinfo{booktitle}{\emph{Proceedings of the 2019 27th ACM Joint Meeting on European Software Engineering Conference and Symposium on the Foundations of Software Engineering}}. \bibinfo{pages}{807--817}.
\newblock


\bibitem[Zhang et~al\mbox{.}(2023)]%
        {where_zhang2023deeplog}
\bibfield{author}{\bibinfo{person}{Yang Zhang}, \bibinfo{person}{Xiaosong Chang}, \bibinfo{person}{Lining Fang}, {and} \bibinfo{person}{Yifan Lu}.} \bibinfo{year}{2023}\natexlab{}.
\newblock \showarticletitle{DeepLog: Deep-Learning-Based Log Recommendation}. In \bibinfo{booktitle}{\emph{2023 IEEE/ACM 45th International Conference on Software Engineering: Companion Proceedings (ICSE-Companion)}}. IEEE, \bibinfo{pages}{88--92}.
\newblock


\bibitem[Zhou et~al\mbox{.}(2023)]%
        {logging_zhou2023scalable}
\bibfield{author}{\bibinfo{person}{Huan Zhou}, \bibinfo{person}{Weining Qian}, \bibinfo{person}{Xuan Zhou}, \bibinfo{person}{Qiwen Dong}, \bibinfo{person}{Aoying Zhou}, {and} \bibinfo{person}{Wenrong Tan}.} \bibinfo{year}{2023}\natexlab{}.
\newblock \showarticletitle{Scalable and adaptive log manager in distributed systems}.
\newblock \bibinfo{journal}{\emph{Frontiers of Computer Science}} \bibinfo{volume}{17}, \bibinfo{number}{2} (\bibinfo{year}{2023}), \bibinfo{pages}{172205}.
\newblock


\bibitem[Zhou et~al\mbox{.}(2019)]%
        {diagnosis_zhou2019latent}
\bibfield{author}{\bibinfo{person}{Xiang Zhou}, \bibinfo{person}{Xin Peng}, \bibinfo{person}{Tao Xie}, \bibinfo{person}{Jun Sun}, \bibinfo{person}{Chao Ji}, \bibinfo{person}{Dewei Liu}, \bibinfo{person}{Qilin Xiang}, {and} \bibinfo{person}{Chuan He}.} \bibinfo{year}{2019}\natexlab{}.
\newblock \showarticletitle{Latent error prediction and fault localization for microservice applications by learning from system trace logs}. In \bibinfo{booktitle}{\emph{Proceedings of the 2019 27th ACM Joint Meeting on European Software Engineering Conference and Symposium on the Foundations of Software Engineering}}. \bibinfo{pages}{683--694}.
\newblock


\bibitem[Zhu et~al\mbox{.}(2015)]%
        {where_zhu2015learning}
\bibfield{author}{\bibinfo{person}{Jieming Zhu}, \bibinfo{person}{Pinjia He}, \bibinfo{person}{Qiang Fu}, \bibinfo{person}{Hongyu Zhang}, \bibinfo{person}{Michael~R Lyu}, {and} \bibinfo{person}{Dongmei Zhang}.} \bibinfo{year}{2015}\natexlab{}.
\newblock \showarticletitle{Learning to log: Helping developers make informed logging decisions}. In \bibinfo{booktitle}{\emph{2015 IEEE/ACM 37th IEEE International Conference on Software Engineering}}, Vol.~\bibinfo{volume}{1}. IEEE, \bibinfo{pages}{415--425}.
\newblock


\end{thebibliography}
\end{document}